\documentclass[journal]{IEEEtran}
\ifCLASSINFOpdf
\else
\fi
\usepackage[utf8]{inputenc}
\usepackage{amsmath,amssymb,amsfonts}
\usepackage{cite}

\usepackage{algorithmic}
\usepackage{stfloats}
\usepackage{graphicx}
\usepackage{textcomp}
\usepackage{xcolor}
\usepackage{color}
\usepackage{pifont}
\usepackage{amsthm}
\newtheorem{theorem}{Theorem}

\usepackage{hyperref}
\usepackage{mathrsfs}
\usepackage{booktabs}
\usepackage{hyperref}
\usepackage{cases}
\usepackage{comment}
\usepackage{empheq} 
\usepackage{caption}
\usepackage{subcaption}
\usepackage{extarrows}
\allowdisplaybreaks[4]
\usepackage{url}  
\usepackage{graphicx}  
\usepackage[ruled,vlined,linesnumbered]{algorithm2e}

\usepackage{listings}

\begin{document}
\captionsetup[figure]{labelfont={rm},labelformat={default},labelsep=period,name={Fig.}}
\title{Distributed Distortion-Aware Robust Optimization for Movable Antenna-aided Cell-Free ISAC Systems}
%
%

\author{Yue Xiu,~\IEEEmembership{Member,~IEEE},~Ning Wei,~\IEEEmembership{Member,~IEEE},~Ran Yang,\\~Zheng Dong,~\IEEEmembership{Member,~IEEE},~Huimin Tang, ~Guangyi Liu,~\IEEEmembership{Member,~IEEE},
\vspace{-3mm}\\
\thanks{Yue Xiu, Ning Wei and Ran yang  are with 
National Key Laboratory of Science and Technology on Communications, University of Electronic Science and Technology of China, Chengdu 611731, China (E-mail:  
xiuyue12345678@163.com, yangran6710@outlook.com, wn@uestc.edu.cn). Zheng Dong is with the School of Information Science and Engineering,
Shandong University, Qingdao 266237, China (e-mail: zhengdong@sdu.edu.cn).
Huimin Tang is with the School of Transportation and Logistics, Southwest Jiaotong University (e-mail: tanghuimin@my.swjtu.edu.cn).
G. Liu are with China Mobile Research
Institute, Beijing 100053, China (email: liuguangyi@chinamobile.com).}
}

\maketitle
\begin{abstract}
The cell-free integrated sensing and communication (CF-ISAC) architecture has emerged as a key enabling technology for future 6G networks, offering efficient spectrum sharing and ubiquitous coverage. However, practical deployments are inevitably subject to hardware impairments, particularly nonlinear distortion caused by power amplifiers (PAs). This can severely degrade both communication rate and sensing accuracy. In this paper, we propose a novel movable antenna (MA)-aided CF-ISAC architecture to mitigate distortion effects and improve system robustness. To account for the nonlinear distortion, we employ a third-order memoryless polynomial distortion to approximate the PAs’ nonlinear distortion. However, in this model, the third-order distortion coefficients (3RDCs) may differ across access points (APs) due to inherent hardware discrepancies, such as variations in PA linearity, device aging, and manufacturing tolerances, as well as environmental and operational factors, including temperature fluctuations. Therefore, we propose a distributed distortion-aware worst-case robust optimization framework that explicitly accounts for the uncertainty caused by the 3RDCs, thereby enhancing the system’s resilience to PAs’ nonlinear distortion. Specifically, we first propose a worst-case analysis to characterize the impact of PA-induced distortion on the Cramer-Rao lower bound (CRLB) and communication rate. Then, to handle the non-convexity and complexity of the worst-case problem, we adopt the successive convex approximation (SCA) algorithm for determining the 3RDCs. Finally, based on the given 3RDCs, we jointly optimize the beamforming and the MA positions of all APs, subject to constraints on transmit power, sensing accuracy. To handle the non-convexity and complexity of the formulated problem, we propose an MA-enabled self-attention and convolutional graph neural network (SACGNN) algorithm for solving the non-convex optimization problem. Simulation results demonstrate that the proposed method significantly improves the communication-sensing trade-off under distortion, and outperforms fixed position antennas (FPAs) baselines in both robustness and system capacity, showing the benefits of MA-aided CF-ISAC systems.
\end{abstract}

\begin{IEEEkeywords}
Cell-free integrated sensing and communication, third-order distortion coefficients, movable antenna. 
\end{IEEEkeywords}

\section{Introduction}
Integrated sensing and communication (ISAC) has emerged as a foundational paradigm in future 6G wireless networks, enabling the joint operation of communication and sensing functionalities over shared spectrum and hardware resources. This integration facilitates enhanced spectral efficiency and reduces hardware cost, making ISAC a key enabled technology for emerging applications such as autonomous vehicles, smart manufacturing, and intelligent surveillance\cite{9737357}. Meanwhile, the cell-free (CF) network architecture has obtained increasing attention due to its potential to eliminate inter-cell interference and improve communication service quality. In CF systems, multiple access points (APs) are densely deployed across a geographical area and centrally coordinated by a central processing unit (CPU), enabling cooperative transmission for user equipments (UEs)\cite{9743355}. The convergence of these two technologies leads to the cell-free integrated sensing and communication (CF-ISAC) system, which leverages distributed APs to simultaneously support high-speed communication and high-precision sensing. However, the practical implementation of CF-ISAC faces several challenges, particularly the impact of hardware impairments such as nonlinear distortion introduced by power amplifiers (PAs). These distortions are often difficult to precisely obtain and can severely degrade the performance of both the communication transmit rate and sensing accuracy. In addition, traditional fixed-position antennas (FPAs) deployment in traditional CF-ISAC systems may limit spatial adaptability. To address these limitations, the use of MAs (MAs) has been proposed to introduce spatial reconfigurability, enabling more flexible and efficient optimization of system resources\cite{10286328}.

\textbf{Challenges.} In practical CF-ISAC systems, nonlinear distortion introduced by PAs becomes a significant bottleneck that limits system performance, particularly under high transmit power conditions. These impairments are typically characterized by third-order nonlinearities, which are difficult to model precisely due to device variability and operating condition fluctuations. Although existing compensation techniques can mitigate distortion in conventional communication systems, they often fail to account for the dual-functional nature of ISAC, where both communication throughput and sensing accuracy are jointly affected. Moreover, uncertainty in the distortion parameters can further exacerbate the problem, leading to substantial degradation in achievable rates and target sensing accuracy. These challenges highlight the necessity of developing a distortion-aware robust optimization framework for CF-ISAC systems.

\textbf{Motivation.} In practical CF-ISAC systems, nonlinear hardware impairments, especially those introduced by PA nonlinearities, significantly deteriorate system performance. These
impairments are commonly modeled by using a Taylor series expansion, where the third-order nonlinear distortion coefficients (3RDCs) capture the dominant intermodulation distortion introduced by the PA. However, in real-world implementations, the value of 3RDCs is often unknown due to multiple
factors:
\begin{itemize}
    \item  Manufacturing inconsistencies across different PAs\cite{9104787}.
    \item  Temperature-dependent behavior and aging effects that cause drift in PA characteristics over time\cite{6954520}.
    \item Dynamic operating conditions, such as changes in output power levels or load impedance\cite{7432149}.
\end{itemize}
Moreover, the challenge of 3RDCs’ uncertainty is further exacerbated in FPAs, where the system cannot adjust its
spatial DoF to mitigate distortion-sensitive regions in the environment. To address these challenges, we propose an MAaided CF-ISAC framework, which introduces additional spatial DoF that can be leveraged to dynamically adjust the MA
positions in response to the PA-induced distortion characteristics and channel conditions. By enabling geometry-aware
reconfiguration, the MA architecture offers new opportunities to spatially decouple the sensing and communication functionalities from distortion-dominated directions. To exploit the full potential of MA-enabled spatial adaptability, we develop a robust distortion-aware optimization framework to ensure reliable joint communication and sensing performance even under 3RDCs’ uncertainty.

The main contributions of this work are summarized as
follows.
\begin{itemize}
    \item First, we consider an MA-aided CF-ISAC system, where distributed APs equipped with MAs jointly serve communication and sensing users over shared time-frequency resources. Under the assumption that the 3RDCs of each AP's PA lie within an unknown bounded interval, we incorporate this uncertainty into the system model to quantify the impact of nonlinear hardware impairments on signal quality. By adopting the worst-case robustness principle, the triangle inequality, and the successive convex approximation (SCA) method, we derive a tractable approximation of the distortion-aware Cramer-Rao lower bound (CRLB) constraint. We demonstrate that the approximation can achieve reliable performance across a wide range of distortions caused by PAs.
    \item After the distortion uncertainty bound of 3RDCs is established, we formulate a distortion-aware joint beamforming and MA positions optimization problem enhance the robustness of the CF-ISAC system. Since the original problem is intractable and involves coupled variables across multiple APs, we propose a novel learning-based solution framework to obtain a high-quality suboptimal policy. Specifically, we model the complex joint optimization as a multi-agent Markov decision process (MDP), where each AP acts as an intelligent agent. An MA-enabled self-attention and convolutional graph neural network (SACGNN) algorithm to learn decentralized policies that coordinate beamforming vectors and MA positions based on shared feedback.  
    \item We propose the robust learning-based optimization algorithm for uncertainty caused by the 3RDCs in the CF-ISAC system, and we analyze its convergence behaviour and computational complexity. Afterwards, the impact of we validate the distortion uncertainty bound of 3RDCs through simulations, and evalute the performance of the proposed algorithm under various levels of nonlinear distortion. Compared with the conventional non-robust designs, the proposed algorithm demonstrates strong robustness against PA-induced distortion uncertainty. Compared with existing robust optimization methods based on convex approximation or deterministic bounding techniques, the proposed SACGNN algorithm achieves significantly better performance in terms of communication quality. Furthermore, the learning-based scheme can flexibly adapt to dynamic system conditions and arbitrary PA distortion levels. 
\end{itemize}
\section{Related Work}

\subsection{PA Nonlinearity Distortion}
Nonlinear distortion introduced by PAs significantly affected the performance of 5G wireless systems, particularly in ISAC contexts. This distortion was typically modeled using Taylor or Volterra series expansions, where the 3RDCs captured the dominant nonlinear distortion effects. The authors in\cite{10646258} conducted a comprehensive study on the robustness of ISAC waveforms under PA-induced nonlinear distortion, demonstrating that the 3RDC played a critical role in degrading sensing accuracy and highlighting the necessity of incorporating nonlinear hardware effects into ISAC system design. The authors in\cite{1255683} analyzed PA distortion products via a detailed Volterra series model and derived closed-form expressions for 3RDC-related distortion, thereby advancing the theoretical understanding of PA nonlinearities and their influence on system linearity and spectral regrowth. The authors in\cite{599544} proposed a Volterra series-based nonlinear PA model that emphasized the significance of 3RDC in PA behavior. The authors in\cite{4717214} explored the impact of third-order nonlinearities on overall communication system performance, focusing on signal degradation caused by 3RDC-related distortions and reinforcing the importance of accurate nonlinear modeling. To mitigate PA nonlinearity, The authors in\cite{1703853} developed generalized memory polynomial (GMP) models for digital predistortion (DPD), effectively capturing the nonlinear memory effects associated with 3RDC and enabling enhanced distortion compensation. Unfortunately, none of the above works considered the inherent uncertainty in 3RDC arising from hardware variability and temperature drift within the CF-ISAC systems employing MAs. This critical gap motivated our study, in which we model 3RDC as an uncertain but bounded parameter and develop a distributed distortion-aware robust optimization framework that jointly optimized beamforming and MA positions to mitigate PA-induced distortion under practical uncertainty.

\subsection{Movable Antenna}
MAs and fluid antennas (FAs) have recently attracted significant attention as innovative techniques for enhancing spatial degrees of freedom and adaptability in wireless communication systems. By dynamically repositioning antenna elements within predefined spatial regions, these techniques enabled flexible channel reconfiguration, improved spatial diversity, and more effective interference management. In \cite{10146262}, Zhang et al. proposed the concept of fluid antenna systems, utilizing fluid-metallic structures to physically adjust antenna positions in real time. Their work thoroughly analyzed the channel variations induced by antenna movement and demonstrated substantial improvements in channel capacity and diversity gain over conventional fixed antenna systems. \cite{10243545} investigated distributed movable antenna systems in cell-free networks, showing that spatial repositioning of antennas at distributed access points improved coverage uniformity and spectral efficiency. They provided theoretical capacity analysis and discussed associated practical implementation challenges. \cite{10978811} explored the use of MAs in near-field ISAC systems. Their study focused on joint waveform design and antenna positioning to simultaneously enhance multi-target sensing accuracy and multi-user communication performance. Simulation results confirmed that MAs effectively leveraged near-field effects for improved ISAC functionality. In the context of cell-free massive MIMO, \cite{11018493} laid the foundational work on distributed architectures without explicit cell boundaries, providing uniformly strong service across users. Their seminal paper addressed key issues such as channel estimation, pilot contamination, and system scalability, establishing a baseline for subsequent MA and FA applications. Despite these advancements, most of the above works assumed ideal hardware and did not account for nonlinear impairments such as PA distortion.

\section{System Model}

\subsection{Transmit Signal Model}
\begin{figure}[!t]
  \centering
  \includegraphics[scale=0.3]{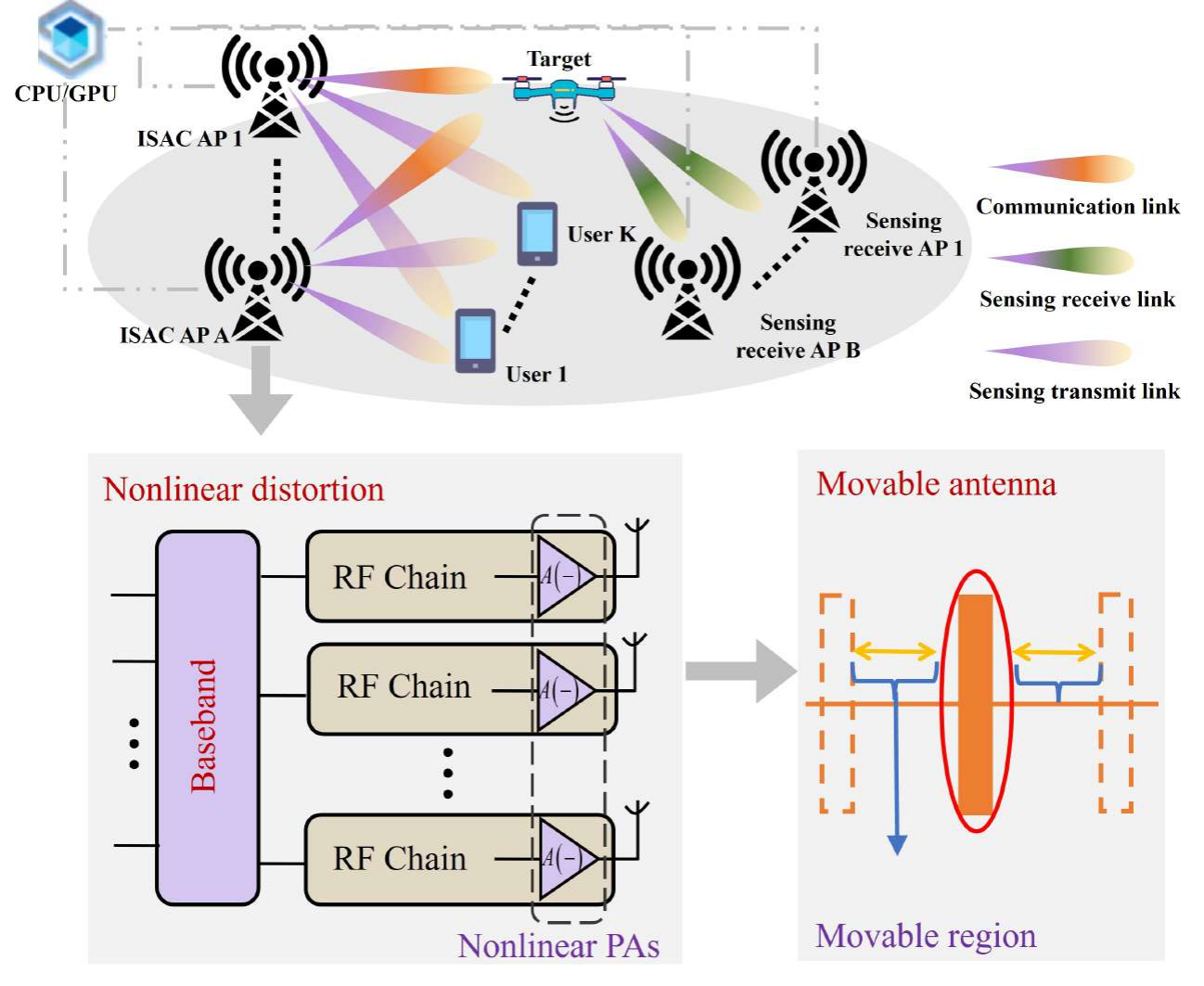}
  \captionsetup{justification=centering}
  \caption{Illustration of the MA-aided CF-ISAC system with nonlinear distortion
.}\vspace{-10pt}
\label{FIGUREPA0}
\end{figure}
As illustrated in Fig.~\ref{FIGUREPA0}, we consider a CF-ISAC system, where each transmit access point (tAP) is equipped with a uniform linear array (ULA) of MAs. Specifically, the system comprises $A$ tAPs and $B$ sensing receive APs (sAPs), each collaboratively serving $K$ single-antenna UEs while sensing a common target over shared time–frequency resources\cite{8606437}. Each tAP $a\in\mathcal{A}$ is equipped with $N_{T}$ MAs, where $\mathcal{A}=\{1,2,\ldots, A\}$, and each is connected to an RF chain and PA. Similarly, each sAP  
$b\in\mathcal{B}$, uses $N_{R}$ receive MAs for sensing, where $\mathcal{B}=\{1,2,\ldots,B\}$. The transmit symbol vector is denoted as $\mathbf{x}=[x_{1},x_{2},\ldots,x_{K}]^{T}\in\mathbb{C}^{K}$, with $\mathbb{E}\{\mathbf{x}\mathbf{x}^{H}\}=\mathbf{I}_{K}$. tAP $a$ applies a beamforming matrix 
$\mathbf{W}_{a}=[\mathbf{w}_{a,1},\ldots,\mathbf{w}_{a,K}]\in\mathbb{C}^{N_{T}\times K}$ to generate the transmitted signal $\mathbf{s}_{a}=\mathbf{W}_{a}\mathbf{x}\in\mathbb{C}^{N_{T}}$, which is then upconverted and amplified by the per–antenna PAs before radiation.

To explicitly capture PA-induced beam distortion, we adopt a third-order memoryless polynomial model \cite{4717214}, and it is expressed as
\begin{align}
z_{a,n}=\mathcal{D}(x_{a,n})=\beta_{1,a}x_{a,n}+\beta_{3,a}x_{a,n}|x_{a,n}|^{2},\forall~a,n,\label{pro1}
\end{align}
where $\mathcal{D}(\cdot)$ is the polynomial operation function. $\beta_{1,a}$ and $\beta_{3,a}\in\mathbb{C}$ express amplitude-to-amplitude modulation (AM/AM) and amplitude-to-phase modulation (AM/PM) distortion effects, respectively. 
$x_{a,n}$ and $z_{a,n}$ denote the $n$-th element of $\mathbf{s}_{a}$ and the PA output, respectively. Recognizing the practical uncertainty in the third-order distortion coefficient due to hardware variations, calibration mismatch, temperature drift, aging, and operating-condition fluctuations, we model the 3RDC at the $a$-th tAP as a nominal value plus an additive perturbation, i.e.,
\begin{align}
\beta_{3,a}=\bar{\beta}_{3,a}+\Delta\beta_{3,a}, \label{pro_2}
\end{align}
where $\bar{\beta}_{3,a}$ denotes the nominal (or estimated) third-order distortion coefficient, and 
$\Delta\beta_{3,a}$ represents the corresponding uncertainty term. We assume that the perturbation is bounded as
\begin{align}
&|\Delta\beta_{3,a}|\leq\epsilon, \label{pro2}
\end{align}
where $\epsilon$ denotes the uncertainty radius. Accordingly, the uncertainty set is defined as
\begin{align}
&\mathcal{E}=\{\beta_{3,a}|\bar{\beta}_{3,a}+\Delta\beta_{3,a},|\Delta\beta_{3,a}|\leq\epsilon,\forall a\}, \label{pro__2}
\end{align}
Specifically, it captures the maximum deviation in $\beta_{3,a}$ caused by hardware impairments, calibration mismatch, or environmental inconsistencies among tAPs. The parameter $\epsilon$ is treated as an uncertain variable in the next proposed robust optimization framework. To facilitate tractable analysis and robust beamforming design, we apply the Bussgang decomposition \cite{9307295} under Gaussian input assumptions. Therefore, we have
\begin{align}
\mathbf{c}_{a}=\boldsymbol{\Xi}_{a}\mathbf{s}_{a}+\mathbf{d}_{a},~\forall~a. \label{pro3}
\end{align}
In this decomposition, $\boldsymbol{\Xi}_{a}\in\mathbb{C}^{N_{T}\times N_{T}}$ denotes the Bussgang gain matrix, which characterizes the linear amplification applied to the input signal, while the vector $\mathbf{d}_{a}=[d_{a,1},\ldots,d_{a,N_{T}}]^{T}\in\mathbb{C}^{N_{T}}$ captures the nonlinear distortion components across the $N_{T}$ MAs. The Bussgang gain matrix 
$\boldsymbol{\Xi}_{a}$ plays a critical role in the modeling process and is defined as
\begin{align}
\boldsymbol{\Xi}_{a}=\tilde{\boldsymbol{\Xi}}_{a}\bar{\boldsymbol{\Xi}}_{a}^{-1},\label{pro4}
\end{align}
Here, (\ref{pro4}) is introduced only to formally define the Bussgang gain matrix in the standard Bussgang form, and it serves as an intermediate step linking the general decomposition in (\ref{pro3}) to the closed-form expression in (\ref{pro5}) for the adopted third-order memoryless polynomial PA model. $\tilde{\boldsymbol{\Xi}}_{a}=\mathbb{E}\{\mathbf{c}_{a}\mathbf{s}_{a}^{H}\}$ denotes the cross-correlation matrix between the nonlinear output $\mathbf{c}_{a}$ and the input signal 
$\mathbf{s}_{a}$, and $\bar{\boldsymbol{\Xi}}_{a}=\mathbb{E}\{\mathbf{s}_{a}\mathbf{s}_{a}^{H}\}$ represents the autocorrelation matrix of the input signal. For the adopted third-order memoryless polynomial PA model in (\ref{pro1}), the Bussgang gain matrix defined in (\ref{pro4}) can be further written in the following closed form:
\begin{align}
\boldsymbol{\Xi}_{a}=\tilde{\boldsymbol{\Xi}}_{a}\bar{\boldsymbol{\Xi}}_{a}^{-1}=\beta_{1}\mathbf{I}_{N_{T}}+2\beta_{3,a}\mathrm{diag}\{\mathbf{W}_{a}\mathbf{W}_{a}^{H}\},\forall~a.\label{pro5}
\end{align}
The distortion vector $\mathbf{d}_{a}\sim\mathcal{CN}(0,\hat{\boldsymbol{\Xi}}_{a})$ is uncorrelated with $\mathbf{s}_{a}$, satisfying $\mathbb{E}\{\mathbf{d}_{a}\mathbf{s}_{a}^{H}\}=\mathbf{0}$\cite{10869384}, and its variance is given by
\begin{align}
\hat{\boldsymbol{\Xi}}_{a}=\mathbb{E}\{\mathbf{d}_{a}\mathbf{d}_{a}^{H}\}=2|\beta_{3,a}|^{2}(\mathbf{W}_{a}\mathbf{W}_{a}^{H}\odot|\mathbf{W}_{a}\mathbf{W}_{a}^{H}|^{2}).\label{pro6}
\end{align}
In the sequel, the worst-case robust design is developed with respect to the uncertainty set in (4), i.e., the perturbation $\Delta \beta_{3,a}$ around the nominal coefficient $\bar{\beta}_{3,a}$, rather than directly imposing a bound on $\beta_{3,a}$ itself.
We further assume that distortion components across different tAPs are statistically independent.

\subsection{MA Channel Model}
To characterize the wireless propagation environment in the considered MA-aided CF-ISAC system, we adopt a geometric multipath channel model that captures the position-dependent spatial characteristics caused by the MAs. In the downlink communication from the $a$-th tAP to the $k$-th single-antenna UE, the channel comprises $L_{a,k}$ propagation paths, each characterized by a complex gain, propagation delay, and angle of departure (AoD). Accordingly, the channel is modeled as follows: \cite{10032173},
\begin{align}
\mathbf{h}_{k}(\mathbf{p}_{a}) = \sum_{l_{a,k}=1}^{L_{a,k}} \alpha_{l_{a,k}} e^{-j 2\pi f_s \tau_{l_{a,k}}} \boldsymbol{\zeta}_{a}(\mathbf{p}_{a}),\label{pro7}
\end{align}
where $\alpha_{l_{a,k}}$ denotes the complex channel coefficient for the $l_{a,k}$-th path, $\tau_{l_{a,k}}$ represents the corresponding path delay, and $f_s$ is the subcarrier frequency. $\mathbf{g}_a(\mathbf{p}_a)$ denotes the field response vector of the MA ULA at AP $a$, given by
\begin{align}
\boldsymbol{\zeta}_{a}(\mathbf{p}_{a}) = \left[1, e^{-j 2\pi p_1^a \sin(\phi_{l_{a,k}})}, \ldots, e^{-j 2\pi p_{N_T}^a \sin(\phi_{l_{a,k}})} \right]^T,\label{pro8}
\end{align}
where $\mathbf{p}_{a} = [p_1^a, \ldots, p_{N_T}^a]^T$ represents the instantaneous positions of the $N_T$ transmit MAs at tAP $a$, and each $p_t^a \in \mathcal{C}_a = [p_a^{\min}, p_a^{\max}]$ defines the feasible movement range of the $t$-th MA. Similarly, for the sensing process between the $a$-th tAP and the $b$-th sAP via a common point target, we adopt a bistatic sensing model following the structure in \cite{7961152}. The sensing channel is given by
\begin{align}
\mathbf{H}_{a,b} = \zeta_{a,b} e^{-j 2\pi f_s \tau_{a,b}} \mathbf{g}_b(\mathbf{p}_b) \bar{\mathbf{g}}_a^H(\mathbf{p}_a),\label{pro9}
\end{align}
where $\zeta_{a,b}$ is the radar cross section (RCS) from the target, and $\tau_{a,b}$ is the round-trip delay from tAP $a$ to the target and then to the sAP $b$, defined as
\begin{align}
\tau_{a,b} = \tau_{a,d} + \tau_{b,d} = \frac{l_{a,d}}{c} + \frac{l_{b,d}}{c},\label{pro10}
\end{align}
where $c$ denotes the speed of light. The distances $l_{a,d}$ and $l_{b,d}$ from the tAPs to the target are calculated as
\begin{align}
&l_{a,d}=\sqrt{(d_x^a-d_x)^2 + (d_y^a-d_y)^2}, \\
&l_{b,d}=\sqrt{(d_x^b-d_x)^2 + (d_y^b-d_y)^2}, \label{pro11}
\end{align}
where $(d_x^a, d_y^a)$ and $(d_x^b, d_y^b)$ are the coordinates of the $a$-th tAP and $b$-th sAP, respectively, and $(d_x, d_y)$ is the coordinate of the target. The field response vectors $\mathbf{g}_b(\mathbf{p}_b)$ and $\bar{\mathbf{g}}_a(\mathbf{p}_a)$ of the receive MA at sAP $b$ and tAP $a$ is defined as
\begin{align}
\mathbf{g}_b(\mathbf{p}_{b}) = \left[1, e^{-j 2\pi p_1^b \sin(\phi_b)}, \ldots, e^{-j 2\pi p_{N_R}^b \sin(\phi_b)} \right]^T,\nonumber\\
\bar{\mathbf{g}}_a(\mathbf{p}_{a}) = \left[1, e^{-j 2\pi p_1^a \sin(\phi_a)}, \ldots, e^{-j 2\pi p_{N_T}^a \sin(\phi_a)} \right]^T,\label{pro12}
\end{align}
where $\mathbf{p}_b = [p_1^b, \ldots, p_{N_R}^b]^T$ denotes the spatial configuration of the $N_R$ receive MAs, and $p_r^b \in \mathcal{C}_b = [p_b^{\min}, p_b^{\max}]$ defines the allowed movement interval for the $r$-th receive MA. To enable precise beam alignment and angular estimation, the angles of departure and arrival are computed geometrically based on the locations of the target and tAPs/sAPs as
\begin{align}
&\phi_a=\arctan\left((d_y - d_y^a)/(d_x - d_x^a) \right) + \pi \cdot \mathbf{1}(d_x < d_x^a), \nonumber\\
&\phi_b=\arctan\left((d_y - d_y^b)/(d_x - d_x^b) \right) + \pi \cdot \mathbf{1}(d_x < d_x^b),\label{pro13}
\end{align}
where $\mathbf{1}(\cdot)$ denotes the indicator function. In this work, we assume that the MA arrays are electronically driven, capable of adjusting element positions in real-time via high-speed control logic, as described in \cite{ning2025movable}. These electronically reconfigurable antennas introduce negligible movement latency, making them suitable for joint dynamic optimization in real-time CF-ISAC operations.

\subsection{Communication System Model}
Each tAP applies beamforming to directionally focus its transmitted signal, followed by nonlinear power amplification to meet the transmission power requirement. However, due to hardware limitations, the PAs at each tAP introduce nonlinear distortion, which severely degrades both communication rate and sensing accuracy\cite{9319879,10662951}. After amplification, the signals propagate through a wireless channel characterized by distance-dependent path loss, spatially varying multipath fading, and additive white Gaussian noise (AWGN). Thus, the received signal at the $k$-th UE can be expressed as 
\begin{align}
y_{k}=\sum_{a=1}^{A}\mathbf{h}_{k}^{H}(\mathbf{p}_{a})\mathbf{c}_{a}+n_{k}=\sum_{a=1}^{A}\mathbf{h}_{k}^{H}(\mathbf{p}_{a})(\boldsymbol{\Xi}_{a}\mathbf{s}_{a}+\mathbf{d}_{a})+n_{k},\label{pro14}
\end{align}
where $\mathbf{h}_k(\mathbf{p}_a) \in \mathbb{C}^{N_T}$ represents the channel vector from the $a$-th tAP (located at position $\mathbf{p}_a$) to the $k$-th UE, $\mathbf{c}_a$ denotes the PA-processed transmitted signal, and $n_k \sim \mathcal{CN}(0, \sigma_k^2)$ denotes the AWGN at the UE. The nonlinear signal received at the $k$-th UE can be decomposed into four terms
\begin{align}
&y_{k}=\underbrace{\sum_{a=1}^{A}\mathbf{h}_{k}^{H}(\mathbf{p}_{a})\boldsymbol{\Xi}_{a}\mathbf{w}_{a,k}x_{k}}_{\text{desired signal}}+\underbrace{\sum_{a=1}^{A}\sum_{j\neq k}^{K}\mathbf{h}_{k}^{H}(\mathbf{p}_{a})\boldsymbol{\Xi}_{a}\mathbf{w}_{a,j}x_{j}}_{\text{inter-UE interference}}\nonumber\\
&+\underbrace{\sum_{a=1}^{A}\mathbf{h}_{k}^{H}(\mathbf{p}_{a})\mathbf{d}_{a}}_{\text{PA impairment}}+\underbrace{n_{k}}_{\text{AWGN}},\label{pro15}
\end{align}
where $\mathbf{w}_{a,k}$ is the beamforming vector from tAP $a$ for UE $k$, $\mathbf{G}_a$ denotes the nominal linear PA gain, and $\mathbf{d}_a$ represents the nonlinear distortion component due to the PA nonlinearity at tAP $a$. To evaluate the system performance under such practical impairments, we define the signal-to-interference-plus-noise-and-distortion ratio (SINDR) at the $k$-th UE is given in (\ref{pro16}) at the top of next page.
\begin{figure*}
\begin{align}
\gamma_{k}=\frac{|\sum_{a=1}^{A}\mathbf{h}_{k}^{H}(\mathbf{p}_{a})\boldsymbol{\Xi}_{a}\mathbf{w}_{a,k}|^{2}}{\sum_{j\neq k}^{K}|\sum_{a=1}^{A}\mathbf{h}_{k}^{H}(\mathbf{p}_{a})\boldsymbol{\Xi}_{a}\mathbf{w}_{a,j}|^{2}+\sum_{a=1}^{A}\mathbf{h}_{k}^{H}(\mathbf{p}_{a})\hat{\boldsymbol{\Xi}}_{a}\mathbf{h}_{k}(\mathbf{p}_{a})+\sigma_{k}^{2}},\forall~k,\label{pro16}
\end{align} 
\hrulefill
\end{figure*}
The beamforming can amplify residual distortion or inadvertently increase inter-user interference, thus degrading both communication quality and sensing accuracy. Moreover, in MA-aided CF-ISAC systems, the positions of each tAP’s MAs can be dynamically adjusted, introducing spatial DoFs into the beamforming process. This added flexibility not only enables enhanced angular resolution for sensing but also affects the spatial characteristics of the distortion components due to 3RDC uncertainties. Therefore, in this paper, we study the robust beamforming design in MA-aided CF-ISAC systems that jointly optimize the beamforming and MA positions to suppress the distortion caused by PAs and the multiuser interference.

\subsection{Sensing System Model}
In this section, we focus on estimating the target location vector $\mathbf{e}$ and evaluate the estimation accuracy by employing the CRLB. Specifically, we consider that each sAP collects reflected signals originating from all $A$ tAPs. The received signal at the $b$-th sAP can be expressed as
\begin{align}
&\mathbf{y}_{b}=\sum_{a=1}^{A}\mathbf{H}_{a,b}(\boldsymbol{\Xi}_{a}\mathbf{s}_{a}+\mathbf{d}_{a})+\mathbf{n}_{b},\label{pro17}
\end{align}
where $\mathbf{n}_{b}\sim\mathcal{CN}(0,\boldsymbol{\Upsilon}_{b})$ and $\boldsymbol{\Upsilon}_{b}=\sigma_{b}^{2}\mathbf{I}$. The unknown target location $\mathbf{e}$ is treated as a deterministic but unknown parameter. In the estimation problems, the CRLB serves as a fundamental performance limit, quantifying the minimum achievable estimation error variance for any unbiased estimator of $\mathbf{e}=[d_{x},d_{y}]^{T}$. The CRLB inequality for the mean squared error (MSE) matrix of the estimator $\hat{\mathbf{e}}$ is given by
\begin{align}
&\mathbb{E}_{\mathbf{y}_{b},\mathbf{e}}\{(\hat{\mathbf{e}}-\mathbf{e})(\hat{\mathbf{e}}-\mathbf{e})^{T}\}\succeq\mathbf{J}_{b}^{-1}(\mathbf{e}),\label{pro18}
\end{align}
where $\mathbf{J}_{b}(\mathbf{e})$ denotes the Fisher information matrix (FIM) associated with $\mathbf{e}$. Then, $\mathbf{J}_{b}(\mathbf{e})$ is expressed as
\begin{align}
&\mathbf{J}_{b}(\mathbf{e})=-\mathbb{E}_{\mathbf{y}_{b}|\mathbf{e}}\left(\frac{\partial^{2}\ln f(\mathbf{y}_{b}|\mathbf{e})}{\partial\mathbf{e}\partial\mathbf{e}^{T}}\right),\label{pro19}
\end{align}
where $f(\mathbf{y}_{b}|\mathbf{e})$ denoting the likelihood function of the observation $\mathbf{y}_{b}$ given by
\begin{align}
&f(\mathbf{y}_{b}|\mathbf{e})=\frac{1}{(2\pi|\boldsymbol{\Upsilon}_{b}|)^{L_{r}/2}}\nonumber\\
&e^{\frac{1}{2}(\mathbf{y}_{b}-\sum_{a=1}^{A}\mathbf{H}_{a,b}(\boldsymbol{\Xi}_{a}\mathbf{x}_{a}+\mathbf{d}_{a}))^{T}\boldsymbol{\Upsilon}_{b}^{-1}(\mathbf{y}_{b}-\sum_{a=1}^{A}\mathbf{H}_{a,b}(\boldsymbol{\Xi}_{a}\mathbf{x}_{a}+\mathbf{d}_{a}))}.\label{pro20}
\end{align}
Then, the CRLB matrix corresponding to $b$-th sAP is formulated as
\begin{align}
&\mathbf{CRLB}_{b}(\mathbf{e})=(\mathbf{J}_{b}(\mathbf{e}))^{-1},\label{pro_21}
\end{align}
where 
$\mathbf{J}_{b}(\mathbf{e})$ represents the FIM, and the detail expression is given in \textbf{Appendix~A}. 

\subsection{Problem Formulation}
To effectively mitigate the nonlinear distortion induced by the PA, especially under modeling errors and uncertainties associated with the 3RDC bound, denoted as $\epsilon$, we adopt a worst-case robust optimization strategy. Specifically, assuming that the third-order distortion coefficient satisfies $\beta_{3,a}=\bar{\beta}_{3,a}+\Delta \beta_{3,a}$ with $\Delta \beta_{3,a}$, we consider the most detrimental impact of this bounded perturbation on the achievable total communication rate and accordingly design the system parameters to ensure robust performance under the worst distortion scenario. To this end, we jointly optimize the beamforming matrices $\{\mathbf{W}_{a}\}_{a=1}^{A}$ of all $A$ tAPs, the positions of their MAs $\{\mathbf{p}_{a}\}_{a=1}^{A}$, and the MA positions $\{\mathbf{p}_{b}\}_{b=1}^{B}$ of the $B$ sAPs. The objective is to maximize the worst-case achievable communication rate subject to the transmit power constraints and the localization accuracy requirements characterized by the Cramér–Rao lower bound (CRLB). This robust formulation facilitates the joint optimization of communication and sensing performance under nonlinear hardware impairments. The mathematical formulation of this problem is presented as
\begin{subequations}
\begin{align}
\max_{\genfrac{}{}{0pt}{}{\{\mathbf{W}_{a}\}_{a=1}^{A}}{\{\mathbf{p}_{a}\}_{a=1}^{A},\{\mathbf{p}_{b}\}_{b=1}^{B}}}\min_{\beta_{3,a}}\sum\nolimits_{k=1}^{K}&\log_{2}(1+\gamma_{k}),\label{pro21a}\\
\mbox{s.t.}~
&\|\mathbf{W}_{a}\|_{F}^{2}\leq P_{t},&\label{pro21b}\\
&\mathrm{Tr}\{\mathbf{CRLB}_{b}(\mathbf{e})\}\leq\gamma_{b},&\label{pro21c}\\
&p_{t}^{a}\in\mathcal{C}_{a}, p_{r}^{b}\in\mathcal{C}_{b},&\label{pro21d}\\
&|p_{t}^{a}-p_{t-1}^{a}|\geq D_{0},&\label{pro21e}\\
&|p_{r}^{b}-p_{r-1}^{b}|\geq D_{0}.&\label{pro21f}\\
&|\beta_{3,a}|\leq\epsilon.&\label{pro21_f}
\end{align}\label{pro21}%
\end{subequations}
Constraint (\ref{pro21b}) denotes the maximum transmit power limitation, where $P_t$ denotes the tAP power budget. Constraint (\ref{pro21c}) ensures that the system maintains a guaranteed sensing accuracy, where $\gamma_b$ specifies the minimum localization accuracy requirement for the sensing task. In constraint (\ref{pro21d}), $\mathcal{C}_a$ and $\mathcal{C}_b$ represent the feasible deployment regions for the MAs associated with the tAP and sAP, respectively. Moreover, in the constraints (\ref{pro21e}) and (\ref{pro21f}), the parameter $D_0$ imposes a minimum distance between any pair of MAs to avoid physical collisions and mitigate mutual coupling effects.

\section{Proposed Algorithm}
The objective in~(\ref{pro21a}) is non-convex due to the coupled optimization variables in its log-fractional terms. To facilitate optimization, we adopt fractional programming (FP) and introduce an auxiliary vector $\boldsymbol{\mu} = [\mu_{1}, \mu_{2}, \ldots, \mu_{K}]^{T}$. Applying the Lagrangian dual transform~\cite{9472958,9348933,10035459} yields the equivalent reformulation in~(\ref{pro22}) at the top of the next page:
\begin{figure*}[t]
\begin{align}
&\sum\nolimits_{k=1}^{K}\log_{2}(1+\mu_{k}) - \sum\nolimits_{k=1}^{K}\mu_{k} + \sum\nolimits_{k=1}^{K}\frac{(1+\mu_{k})\left|\sum_{a=1}^{A}\mathbf{h}_{a,k}^{H}\boldsymbol{\Xi}_{a}\mathbf{w}_{a,k}\right|^{2}}{\sum\nolimits_{j=1}^{K}\left|\sum_{a=1}^{A}\mathbf{h}_{a,k}^{H}\boldsymbol{\Xi}_{a}\mathbf{w}_{a,j}\right|^{2} + \sum_{a=1}^{A}\mathbf{h}_{a,k}^{H}\mathbf{C}_{d,a}\mathbf{h}_{a,k} + \sigma_{k}^{2}}, \label{pro22}
\end{align}
\hrulefill
\end{figure*}

The equivalence between~(\ref{pro21a}) and~(\ref{pro22}) holds when the auxiliary variable $\mu_{k}$ is set to its optimal value:
\begin{figure*}[t]
\begin{align}
\mu_{k}^{*} = \frac{\left|\sum\nolimits_{a=1}^{A}\mathbf{h}_{k}^{H}(\mathbf{p}_{a})\boldsymbol{\Xi}_{a}\mathbf{w}_{a,k}\right|^{2}}{\sum\nolimits_{j\neq k}^{K}\left|\sum\nolimits_{a=1}^{A}\mathbf{h}_{k}^{H}(\mathbf{p}_{a})\boldsymbol{\Xi}_{a}\mathbf{w}_{a,j}\right|^{2} + \sum\nolimits_{a=1}^{A}\mathbf{h}_{k}^{H}(\mathbf{p}_{a})\hat{\boldsymbol{\Xi}}_{a}\mathbf{h}_{k}(\mathbf{p}_{a}) + \sigma_{k}^{2}}. \label{pro23}
\end{align}
\hrulefill
\end{figure*}
Here, $\mu_{k}$ is the auxiliary variable introduced by fractional programming, and it is updated according to the current beamforming matrices and MA positions.
Although this reformulation alleviates the logarithmic-fraction coupling, the summation of fractional terms in~(\ref{pro22}) still poses challenges for direct optimization. To further facilitate optimization over $\{\mathbf{W}_{a}\}_{a=1}^{A}$, we apply the quadratic transform technique, which transforms the third term in~(\ref{pro22}) into the following form:
\begin{align}
2\sqrt{1+\mu_{k}}\,\mathrm{Re}\left\{\zeta_{k}^{*}\sum\nolimits_{a=1}^{A}\mathbf{h}_{a,k}^{H}\boldsymbol{\Xi}_{a}\mathbf{w}_{a,k}\right\} - |\zeta_{k}|^{2}\Psi_{k}, \label{pro24}
\end{align}
where, for notational convenience, we define
\begin{align}
&\Psi_{k} = \sum\nolimits_{j=1}^{K}\left|\sum\nolimits_{a=1}^{A}\mathbf{h}_{a,k}^{H}\boldsymbol{\Xi}_{a}\mathbf{w}_{a,j}\right|^{2} + \sum\nolimits_{a=1}^{A}\mathbf{h}_{a,k}^{H}\mathbf{C}_{d,a}\mathbf{h}_{a,k}\nonumber\\
&+\sigma_{k}^{2}. \label{pro25}
\end{align}
Here, $\zeta_{k}$ is the $k$-th element of the auxiliary vector $\boldsymbol{\zeta} = [\zeta_{1}, \zeta_{2}, \ldots, \zeta_{K}]^{T}$. The equivalence between~(\ref{pro22}) and~(\ref{pro24}) holds when $\zeta_{k}$ is set to its optimal value:
\begin{align}
\zeta_{k}^{*} = \sqrt{1+\mu_{k}}\sum\nolimits_{a=1}^{A}\mathbf{h}_{a,k}^{H}\boldsymbol{\Xi}_{a}\mathbf{w}_{a,k}/\Psi_{k}. \label{pro26}
\end{align}
Similarly, $\zeta_{k}$ is the auxiliary variable introduced by the quadratic transform and is updated according to the current optimization variables.
It is worth noting that the auxiliary variables $\mu_{k}$ and $\zeta_{k}$ are not fixed constants throughout the optimization process. Instead, for any given beamforming matrices $\{\mathbf{W}_{a}\}_{a=1}^{A}$ and MA positions $\{\mathbf{p}_{a}\}_{a=1}^{A}$, $\{\mathbf{p}_{b}\}_{b=1}^{B}$, the optimal $\mu_{k}$ and $\zeta_{k}$ are updated according to their closed-form expressions in (\ref{pro23}) and (\ref{pro26}), respectively. Therefore, $\mu_{k}$ and $\zeta_{k}$ should be understood as iteration-dependent auxiliary variables associated with the current values of the optimization variables.
Substituting the reformulated expression~(\ref{pro24}) into the objective, the transformed version in~(\ref{pro21a}) becomes:
\begin{align}
&\sum\nolimits_{k=1}^{K}\left( \log_{2}(1+\mu_{k}) - \mu_{k} + 2\sqrt{1+\mu_{k}}\,\mathrm{Re}\left\{\zeta_{k}^{*}\sum\nolimits_{a=1}^{A}\mathbf{h}_{a,k}^{H}\right.\right.\nonumber\\
&\left.\left.\times\boldsymbol{\Xi}_{a}\mathbf{w}_{a,k}\right\} - |\zeta_{k}|^{2} \Psi_{k} \right). \label{pro27}
\end{align}
For notational compactness and to facilitate subsequent optimization, expression~(\ref{pro27}) is rearranged as:
\begin{align}
\sum\nolimits_{k=1}^{K} \left( \log_{2}(1+\mu_{k}) - \mu_{k} - |\zeta_{k}|^{2} \sigma_{k}^{2} \right) + \delta, \label{pro28}
\end{align}
where $\delta$ is defined as:
\begin{align}
&\delta = \sum\nolimits_{k=1}^{K} \Bigg( 2\sqrt{1+\mu_{k}}\,\mathrm{Re}\left\{\zeta_{k}^{*}\sum\nolimits_{a=1}^{A}\mathbf{h}_{a,k}^{H} \boldsymbol{\Xi}_{a} \mathbf{w}_{a,k} \right\} 
- |\zeta_{k}|^{2}\nonumber\\
&\sum\nolimits_{j=1}^{K} \left|\sum\nolimits_{a=1}^{A} \mathbf{h}_{a,k}^{H} \boldsymbol{\Xi}_{a} \mathbf{w}_{a,j} \right|^{2} - |\zeta_{k}|^{2} \sum\nolimits_{a=1}^{A} \mathbf{h}_{a,k}^{H} \mathbf{C}_{d,a} \mathbf{h}_{a,k} \Bigg). \label{pro29}
\end{align}
In summary, the original objective function in~(\ref{pro21a}) has now been transformed into a more tractable structure via a series of fractional and quadratic transformations. Intuitively, this reformulated objective can be interpreted as the sum of transformed signal-to-interference-plus-distortion ratios (SINDRs) for all $K$ UEs, where the original ratio structure has been decoupled and expressed as a difference between the desired signal power and multiuser interference plus nonlinear distortion. In the following this section, we develop efficient optimization algorithms based on this reformulation to further solve the problem and derive the optimal system design. Specifically, the optimization is carried out in an alternating manner. For given auxiliary variables $\mu_{k}$ and $\zeta_{k}$, the beamforming matrices and MA positions are optimized by solving the transformed problem. Then, based on the updated $\mathbf{W}_{a}$, and $\mathbf{p}_{b}$, the auxiliary variables 
$\mu_{k}$ and $\zeta_{k}$ are refreshed via (\ref{pro23}) and (\ref{pro26}). This alternating update procedure is repeated until convergence.

In the conventional centralized beamforming design, the optimization is typically performed in an iterative manner by alternately updating the auxiliary variables $\boldsymbol{\mu}$ and $\boldsymbol{\zeta}$, the beamforming matrices $\{\mathbf{W}_{a}\}_{a=1}^{A}$, and the positions of the MAs, denoted as $\{\mathbf{p}_{a}\}_{a=1}^{A}$ and $\{\mathbf{p}_{b}\}_{b=1}^{B}$. In practical implementations, the locations of the MAs at the transmitter and receiver need to be properly synchronized to ensure coherent joint beamforming and sensing performance\cite{10286328}.
By observing the equivalent objective function derived in (\ref{pro28}), it is clear that when the auxiliary variables $\boldsymbol{\mu}$ and $\boldsymbol{\zeta}$ are fixed, the resulting objective function for optimizing the beamforming and MA positions reduces to the term $\delta$ defined in (\ref{pro29}), which includes only the relevant components associated with these variables. This simplification enables the original problem to be reformulated as follows:
\begin{subequations}
\begin{align}
\max_{\genfrac{}{}{0pt}{}{\{\mathbf{W}_{a}\}_{a=1}^{A}}{\{\mathbf{p}_{a}\}_{a=1}^{A},\{\mathbf{p}_{b}\}_{b=1}^{B}}} \min_{\beta_{3,a}} & \delta, \label{pro30a}\\
\text{s.t.} \quad & (\ref{pro21b})-(\ref{pro21f}). \label{pro30b}
\end{align} \label{pro30}%
\end{subequations}
However, it is important to note that the closed-form value of the CSI uncertainty bound $\epsilon$ remains undetermined at this stage. To proceed with solving the above problem in (\ref{pro30}), one must first derive an appropriate bound on $\epsilon$ based on the constraint $|\beta_{3,a}| \leq \epsilon$. The detailed derivation of $\epsilon$ will be presented in the following section.

\subsection{Derivation of the 3RDCs bound}
This section is dedicated to the derivation of the 3RDCs bound in (\ref{pro2}) according to $|\beta_{3,a}|\leq\epsilon$, where $\epsilon$ denotes the upper bound of $|\beta_{3,a}|$, which is unknown and should be derived according to $|\beta_{3,a}|\leq\epsilon$\cite{10027476}. It is remarkable that when $a$ are given, deriving $\epsilon$ is equivalent to deriving the minimum of $\delta$ and the maximum of the CRLB under the constraint of $|\beta_{3,a}|\leq\epsilon$. This is because the derived $\delta$ is higher than the practical $\delta$, and the CRLB is smaller than the practical CRLB. The worst-case 3RDCs bound experienced during the optimization process will become even better than the practical worst-case 3RDCs bound, hence resulting in a less robust solution for problem (\ref{pro30})\cite{10027476}. Consequently, in order to preserve the robustness of the transmit and passive beamforming, we need to derive an approximate upper bound of the minimum of $\delta$ under $|\beta_{3,a}|\leq\epsilon$. The result is provided in the following \textbf{Theorem}~\ref{th1}.
\begin{theorem}\label{th1}
When $|\beta_{3,a}|\leq\epsilon$, the worst-case robust optimization problem can be derived as
\begin{subequations}
\begin{align}
\max_{\epsilon}&~\mathcal{L}(\epsilon),\label{pro31a}\\
\mbox{s.t.}~
&c + \epsilon \Theta + \epsilon^2 \dot{\Theta}+2\Pi\leq g(\epsilon,\epsilon_{0})&\label{pro31b}
\end{align}\label{pro31}%
\end{subequations}
where the expressions of $g(\epsilon,\epsilon_{0})$ and $\mathcal{L}(\epsilon)$ are given by
\begin{align}
&\mathcal{L}(\epsilon)=\sum\nolimits_{k=1}^{K}(2\sqrt{1+\mu_{k}}(-\epsilon\sum\nolimits_{a=1}^{A}|\zeta_{k}^{*}2\mathbf{h}_{a,k}^{H}\nonumber\\
&\times\mathrm{diag}\{\mathbf{W}_{a}\mathbf{W}_{a}\}\mathbf{w}_{a,k}|)-|\zeta_{k}|^{2}\sum\nolimits_{j=1}^{K}|\sum\nolimits_{b=1}^{B}\beta_{1}\nonumber\\
&\mathbf{h}_{b,k}^{H}\mathbf{w}_{b,j}|^{2}+\epsilon^{2}|\sum\nolimits_{b=1}^{B}2\mathbf{h}_{b,k}^{H}\mathrm{diag}\{\mathbf{W}_{a}\mathbf{W}_{a}^{H}\}\mathbf{w}_{b,j}|^{2}\nonumber\\
&-\sum\nolimits_{a=1}^{A}2\epsilon^{2}\mathbf{h}_{a,k}^{H}(\mathbf{W}_{a}\mathbf{W}_{a}^{H}\odot|\mathbf{W}_{a}\mathbf{W}_{a}^{H}|^{2})\mathbf{h}_{a,k}),\nonumber\\
&g(\epsilon,\epsilon_{0})=\gamma_b \bigg[ 
- \epsilon (c+2\Pi)(\Theta_1 + \Theta_2) 
+ \epsilon^2 \left( (c+2\Pi)(\dot{\Theta}_1\right.\nonumber\\
&\left.+ \dot{\Theta}_2) + \Theta_1 \Theta_2 \right)-\left(\epsilon^3\right)(\Theta_1 \dot{\Theta}_2 + \Theta_2 \dot{\Theta}_1)+ \left(\epsilon^4 \right)\dot{\Theta}_1 \dot{\Theta}_2\nonumber\\ 
&-\epsilon (c+2\Pi)(\Theta_3+\Theta_4)-\epsilon^2 \left( (c+2\Pi)(\dot{\Theta}_3 + \dot{\Theta}_4) + \Theta_3 \Theta_4 \right)\nonumber\\
&- \left(\epsilon^3 \right)(\Theta_3 \dot{\Theta}_4 + \Theta_4 \dot{\Theta}_3)-\left(\epsilon^4\right) \dot{\Theta}_3 \dot{\Theta}_4 
\bigg].\label{pro32}
\end{align}
This proof is given in \textbf{Appendix}~\textbf{A}. Since the problem in (\ref{pro31}) is convex, we can use CVX\cite{grant2011cvx} to solve this problem.
\end{theorem}

\subsection{Problem Formulation}
Each diagonal element of the CRLB matrix represents the minimum variance of the corresponding parameter estimated by an unbiased estimator. Therefore, the trace of the CRLB matrix is utilized to characterize the performance of target estimation. Our objective is to maximize the CRLB of target estimation in the presence of TS errors by designing the positions of the MAs at the APs and coordinating the transmit beamforming, while satisfying the communication rate and transmit power constraints. Notably, to achieve a robust joint design for the system, we focus on maximizing the worst-case CRLB by optimizing the TS error 
$\bar{\tau}_{a,b}$, ensuring that the proposed design remains resilient to uncertainties arising from varying TS errors. Specifically, the problem of maximizing sensing accuracy, subject to the communication rate constraint, can be formulated as 
\begin{subequations}
\begin{align}
\max_{\genfrac{}{}{0pt}{}{\{\mathbf{W}_{a}\}_{a=1}^{A}}{\{\mathbf{p}_{a}\}_{a=1}^{A},\{\mathbf{p}_{b}\}_{b=1}^{B}}}\min_{\beta_{3,a}}&\delta,\label{pro33a}\\
\mbox{s.t.}~
&(\ref{pro21b})-(\ref{pro21f}).\label{pro33b}
\end{align}\label{pro33}%
\end{subequations} 
This max–min structure complicates the optimization process. To overcome this difficulty, we introduce an auxiliary variable $\varphi$ to represent the worst-case aggregated performance metric (i.e., the minimum of $\sum_k \delta$ over $\beta_{3,a}$), and transform the original problem into the following equivalent form:
\begin{subequations}
\begin{align}
\max_{\{\mathbf{W}_{a}\}, \{\mathbf{p}_{a}\}, \{\mathbf{p}_{b}\}, \kappa}~ 
& \kappa \label{eq:reform_obj} \\
\text{s.t.}~~ 
&\delta \geq \kappa,~\forall~\beta_{3,a} \label{eq:reform_innerbound} \\
&(\ref{pro21b})-(\ref{pro21f}) \label{eq:reform_const}
\end{align}
\label{eq:reform_problem}%
\end{subequations}
In the reformulated problem~\eqref{eq:reform_problem}, constraint~\eqref{eq:reform_innerbound} ensures that the performance metric $\sum_{k=1}^{K} \delta$ under any possible value of $\beta_{3,a}$ remains no smaller than the auxiliary variable $\varphi$, thereby preserving the worst-case behavior captured in the original formulation.

\section{Problem Reformulation for SACGNN}
Due to the coupled nonconvexity and high dimensionality, classical optimization techniques suffer from prohibitive complexity, especially for large-scale systems with many MAs and users. By modeling the problem as a heterogeneous graph $\mathcal{G} = (\mathcal{V}, \mathcal{E})$, the heterogeneous graph $\mathcal{G}$ comprises: tAP nodes, sAP nodes, and UE nodes. $\mathcal{V}_{t}=\{V_{t,1},\ldots,V_{t,A}\}$ represents the set of tAP nodes,  $\mathcal{V}_{s}=\{V_{s,1},\ldots,V_{s,B}\}$ represents the set of sAP nodes, $\mathcal{V}_{u}=\{V_{u,1},\ldots,V_{u,K}\}$ represents the set of UE nodes, $\mathbf{e}_{a,k}$ represents the communication channel feature edges connecting the $a$-th tAP node to  $k$-th UE node. $\tilde{\mathbf{e}}_{a,b}$ represents the sensing channel feature edges connecting $a$-th tAP node to $b$-th rAP node. Then, we have $\mathcal{V}=\mathcal{V}_{t}\bigcup\mathcal{V}_{s}\bigcup\mathcal{V}_{u}$. The initial hidden states $\mathbf{g}^{(1)}[V_{t,a}]$, $\mathbf{g}^{(1)}[V_{s,b}]$ and $\mathbf{g}^{(1)}[V_{u,k}]$ at each node $V_{t,a}$, $V_{s,b}$ and $V_{u,k}$ , respectively, are constructed from the edge features representing channel coefficients, positional information, and hardware distortion parameters.
Using Transformer-based attention mechanisms, the SACGNN enables nodes to selectively aggregate relevant information from heterogeneous neighbors, effectively capturing the interactions among beamforming weights, antenna positions, and hardware distortion effects\cite{9911215}.
The final hidden states after $L$ layers are decoded to yield optimized beamforming matrices ${\mathbf{W}_a}$ and antenna positions ${\mathbf{p}_a, \mathbf{p}_b}$ satisfying the power, CRLB, and mobility constraints. Each node denotes an antenna element or a user, and its initial features include channel statistics, position priors, and hardware distortion information. Each edge models a wireless channel or an interference link via channel gains and path loss. By iterative message passing, the proposed graph learning framework updates node features to optimize beamforming and antenna positions in a distributed manner. As a result, it achieves scalable optimization with improved robustness against uncertainty and time-varying network conditions.

\subsection{Message Passing and Transformer Attention}
The SACGNN employs Transformer-style self-attention to perform heterogeneous message aggregation:
\begin{align}
\mathbf{m}_{v}^{(l)}=\sum_{u\in\mathcal{N}(v)}e_{vu}^{(l)}\mathbf{W}_{v}^{(l)}
\mathbf{h}_{u}^{(l-1)}+\sum_{\tilde{u}\in\tilde{\mathcal{N}}(v)}\tilde{e}_{v\tilde{u}}^{(l)}\mathbf{W}_{v}^{(l)}
\mathbf{h}_{\tilde{u}}^{(l-1)},\label{pro37}
\end{align}
where $v\in\mathcal{V}$, $l\in\{1,\ldots,L\}$, $\mathcal{N}(v)\in\mathcal{V}_{t}\bigcup\mathcal{V}_{u}$,$\tilde{\mathcal{N}}(v)\in\mathcal{V}_{t}\bigcup\mathcal{V}_{s}$. $\mathbf{h}_u^{(l-1)}$ and $\mathbf{h}_{\tilde{u}}^{(l-1)}$ are the hidden feature vectors of neighbor node at layer $l-1$, $\mathbf{W}_v^{(l)}$ is the learnable transformation matrix for node $v$ at layer $l$, $\alpha_{vu}^{(l)}$ and $\alpha_{v\tilde{u}}^{(l)}$ are the attention weight and they are computed as\cite{vaswani2017attention}
\begin{align}
\alpha_{v\bar{u}}=\mathrm{exp}\left((\mathbf{q}_{v}^{(l)})^{T}\mathbf{k}_{\bar{u}}^{(l)}/\sqrt{d_{k}}\right),\label{pro38}
\end{align}
where $\bar{u}\in\{u,\tilde{u}\}$, $\mathbf{q}_V^{(l)}$, and $\mathbf{k}_U^{(l)}$ are query and key vectors, respectively,
and $d_k$ is the dimension scaling factor. Through this mechanism, the nodes selectively integrate messages from neighbors based on learned importance scores, enabling efficient encoding of the coupled interference, channel conditions, and distortion effects. After $L$ layers of message passing, the final node embeddings $\mathbf{h}_V^{(L)}$ are decoded to produce: Beamforming vectors $\mathbf{w}_{a,k}$ for each tAP $a$ and UE $k$, MA position updates $\mathbf{p}_a$, $\mathbf{p}_b$. These outputs are then projected onto feasible sets to satisfy power, CRLB, and MA position constraints, e.g., via normalization or clipping. The SACGNN is trained in an unsupervised manner to maximize the worst-case sum-rate while satisfying sensing and MA position constraints by minimizing a composite loss function
\begin{equation}
   \mathcal{L}
   =
   -\sum_{k=1}^{K}\kappa
   \;-\;
   \lambda\sum_{i=1}^{C_1}\mathbf{1}_{\text{vio},i},
   \quad
   \beta>0.
   \label{pro39}
\end{equation}
where $C_{i}$ is the number of the constraint conditions of problem (\ref{eq:reform_problem}). $\lambda$ is the penalty coefficient. Then, we provide the rigorous mathematical formulation of the Transformer-based SACGNN designed to solve the robust distortion-aware joint beamforming and MA optimization problem in CF-ISAC systems. Each node $V \in \mathcal{V}$ is associated with an initial feature vector $\mathbf{h}_V^{(0)}$. tAP node $V_{a,m}$ (the $m$-th MA at tAP $a$) is defined as
\begin{align}
\mathbf{h}_{V_{a,m}}^{(0)}=[\mathbf{p}_{a},\beta_{3,a}, \mathrm{PC}_{a},\mathbf{c}_{a,m}]\in\mathbb{R}^{N_{T}+4},\label{pro40}
\end{align}
where $\mathbf{p}_a$ is the MA position vector, $\beta_{3,a}$ is the PA distortion coefficient, $\mathrm{PC}_{a}$ is normalized power limit, and $\mathbf{c}_{a,m}$ is a channel-related feature embedding (e.g., channel statistics or covariance eigenvalues) associated with MA $m$. Then, sAP node $V_{b,n}$ (the $n$-th MA at sAP $b$) is expressed as
\begin{align}
\mathbf{h}_{V_{b,n}}^{(0)}=[\mathbf{p}_{b},\gamma_{b},\mathbf{s}_{b,n}]\in\mathbb{R}^{N_{R}+2},\label{pro41}
\end{align}
where $\mathbf{p}_b$ is the MA position, $\gamma_b$ the sensing accuracy threshold, and $\mathbf{s}_{b,n}$ sensing channel feature embedding. UE node $V_k$ is given by
\begin{align}
\mathbf{h}_{V_{k}}^{(0)}=[\mathbf{q}_{k},\sigma_{k}^{2},\mathbf{u}_{k}]\in\mathbb{R}^{5},\label{pro42}
\end{align}
where $\mathbf{q}_k$ is the UE spatial position, $\sigma_k^2$ noise power, and $\mathbf{u}_k$ is communication channel statistical features. For each edge $(U,V) \in \mathcal{E}$, where $U$ and $V$ are connected nodes, we define edge feature embeddings $\mathbf{o}_{U,V}$ to capture: Channel state information (CSI) between MAs or MA-to-UE. Path loss and spatial distance metrics. Hardware distortion correlation features if applicable. These edge features can be initialized by domain knowledge or learned through embedding layers.

\subsection{Layer-wise Message Passing with Transformer Attention}
At each layer $l$, the hidden feature vector $\mathbf{h}_v^{(l)}$ of node $V$ is updated by aggregating information from its neighbors $\mathcal{N}(V)$ via the following Transformer-style self-attention mechanism
\begin{align}
&\mathbf{h}_v^{(l)} = \mathrm{LayerNorm} \left( \mathbf{h}_v^{(l-1)} + \mathrm{MultiHeadAttn} \left( \mathbf{h}_v^{(l-1)}, \right.\right.\nonumber\\
&\left.\left.{\mathbf{h}_{\bar{u}}^{(l-1)} | \bar{u} \in \mathcal{N}(v)}, {\mathbf{o}_{\bar{u},v} |\bar{u} \in \mathcal{N}(v)} \right) \right),\label{pro43}
\end{align}
where $\mathrm{LayerNorm}$ denotes layer normalization to stabilize training, and $\mathrm{MultiHeadAttn}$ is a multi-head attention module adapted for heterogeneous graphs.

\textbf{Multi-Head Attention Computation\cite{vaswani2017attention}:} For each attention head $h = 1,\dots, H$, we compute queries, keys, and values as follows:
\begin{align}
&\mathbf{q}_{v}^{(h)}=\mathbf{W}_{q}^{(l,h)}\mathbf{q}_{v}^{(l-1)},\mathbf{k}_{u}^{(h)}=\mathbf{W}_{k}^{(l,h)}\mathbf{h}_{\bar{u}}^{(l-1)}+\mathbf{W}_{e}^{(l,h)}\mathbf{o}_{\bar{u},v}\nonumber\\
&\mathbf{v}_{\bar{u}}^{(h)}=\mathbf{W}_{v}^{(l,h)}\mathbf{v}_{\bar{u}}^{(l-1)},\label{pro44}
\end{align}
where 
\begin{align}
\mathbf{W}_q^{(l)} = 
\begin{bmatrix}
\mathbf{W}_q^{(1,l)},\ldots,
\mathbf{W}_q^{(H,l)}
\end{bmatrix}^{T}
\in \mathbb{R}^{(H \cdot d_{\text{head}}) \times d_{\text{model}}},\nonumber\\
\mathbf{W}_k^{(l)} = 
\begin{bmatrix}
\mathbf{W}_k^{(1,l)},\ldots,
\mathbf{W}_k^{(H,l)}
\end{bmatrix}^{T}
\in \mathbb{R}^{(H \cdot d_{\text{head}}) \times d_{\text{model}}},\nonumber\\
\mathbf{W}_v^{(l)} = 
\begin{bmatrix}
\mathbf{W}_v^{(1,l)},\ldots,
\mathbf{W}_v^{(H,l)}
\end{bmatrix}^{T}
\in \mathbb{R}^{(H \cdot d_{\text{head}}) \times d_{\text{model}}},\nonumber\\
\mathbf{W}_e^{(l)} = 
\begin{bmatrix}
\mathbf{W}_e^{(1,l)},\ldots,
\mathbf{W}_e^{(H,l)}
\end{bmatrix}^{T}
\in \mathbb{R}^{(H \cdot d_{\text{head}}) \times d_{\text{model}}},\label{pro45}
\end{align}
where $\mathbf{W}_q^{(h,l)}$, $\mathbf{W}_k^{(h,l)}$, $\mathbf{W}_v^{(h,l)}$, and $\mathbf{W}_e^{(h,l)}$ are learnable projection matrices. The $h$-th head attention weights $\alpha_{v\bar{u}}^{(h)}$ are computed by scaled dot-product
\begin{align}
\alpha_{v\bar{u}}^{(h)}=e^{\frac{\mathbf{q}_{v}^{(h)T}\mathbf{k}_{\bar{u}}^{(h)}}{\sqrt{d_{k}}}}/\sum\nolimits_{\bar{u}\in\mathcal{N}(V)} e^{\frac{\mathbf{q}_{v}^{(h)T}\mathbf{k}_{\bar{u}}^{(h)}}{\sqrt{d_{k}}}},\label{pro46}
\end{align}
The aggregated message for head $h$ is
\begin{align}
\mathbf{m}_{v}^{(h)}=\sum_{\bar{u}\in\mathcal{N}(v)}\alpha_{v\bar{u}}^{(h)}\mathbf{v}_{\bar{u}}^{(h)}.\label{pro47}
\end{align}
The multi-head output concatenates all head outputs
\begin{align}
\mathrm{MultiHeadAttn}(\cdot)=\mathbf{W}_{O}\mathrm{Concat}(\mathbf{m}_{v}^{(1)},\ldots,\mathbf{m}_{v}^{(H)}),\label{pro48}
\end{align}
where $\mathbf{W}_O$ is a learnable output projection matrix.

\textbf{Feed-Forward Network (FFN)\cite{vaswani2017attention}:} Optionally, after attention, a position-wise FFN is applied:
\begin{align}
\mathbf{h}_{v}^{(l)}=\mathrm{LayerNorm}(\mathbf{h}_{v}^{(l)}+\mathrm{FFN}(\mathbf{h}_{v}^{(l)})),\label{pro49}
\end{align}
where
\begin{align}
\mathrm{FFN}(\mathbf{x})=\mathrm{ReLU}(\mathbf{x}\mathbf{W}_{1}+\mathbf{b}_{1})\mathbf{W}_{2}+\mathbf{b}_{2},\label{pro50}
\end{align}
with $\mathbf{W}_1, \mathbf{W}_2, \mathbf{b}_1$, and $\mathbf{b}_2$ as learnable parameters.

\subsection{Decoding Optimized Variables from Node Embeddings}
After $L$ layers, the final node embeddings $\mathbf{h}_v^{(L)}$ are decoded to obtain: For each tAP node $v_{a,m}$, the beamforming coefficients $\mathbf{w}_{a,k}(m)$:
\begin{align}
\hat{\mathbf{w}}_{a,k}(m)=f_{w}(\mathbf{h}_{v_{a,m}}^{(L)};\theta_{w}),\label{pro51}
\end{align}
where $f_w(\cdot)$ is a small multi-layer perceptron (MLP) with parameters $\theta_w$. For each tAP $a$ and rAP $b$, the updated antenna positions:
\begin{align}
\hat{\mathbf{p}}_{a}=f_{p}(\{\mathbf{h}_{v_{a,m}}^{(L)}\}_{m=1}^{M};\theta_{p}),\hat{\mathbf{p}}_{b}=f_{p}(\{\mathbf{h}_{v_{b,n}}^{(L)}\}_{n=1}^{M};\theta_{p}),\label{pro52}
\end{align}
where $f_p(\cdot)$ aggregates the antenna element embeddings via pooling (e.g., mean or max) followed by an MLP with parameters $\theta_p$.
To ensure constraints, outputs are projected:
\begin{align}
&\mathbf{p}_{a}=\mathrm{Proj}_{\mathcal{C}_{t}}(\hat{\mathbf{p}}_{a}), \mathbf{p}_{b}=\mathrm{Proj}_{\mathcal{C}_{r}}(\hat{\mathbf{p}}_{b}),\nonumber\\
&\mathbf{W}_{a}=\mathrm{Proj}_{\|\cdot\|_{F}}(\hat{\mathbf{W}}_{a}).\label{pro53}
\end{align}
The projection operators clip or normalize the outputs to meet power budgets and spatial feasibility

\subsection{Loss Function and Training Objective}
\begin{figure}[htbp]
  \centering
  \includegraphics[width=0.5\textwidth, height=0.25\textwidth]{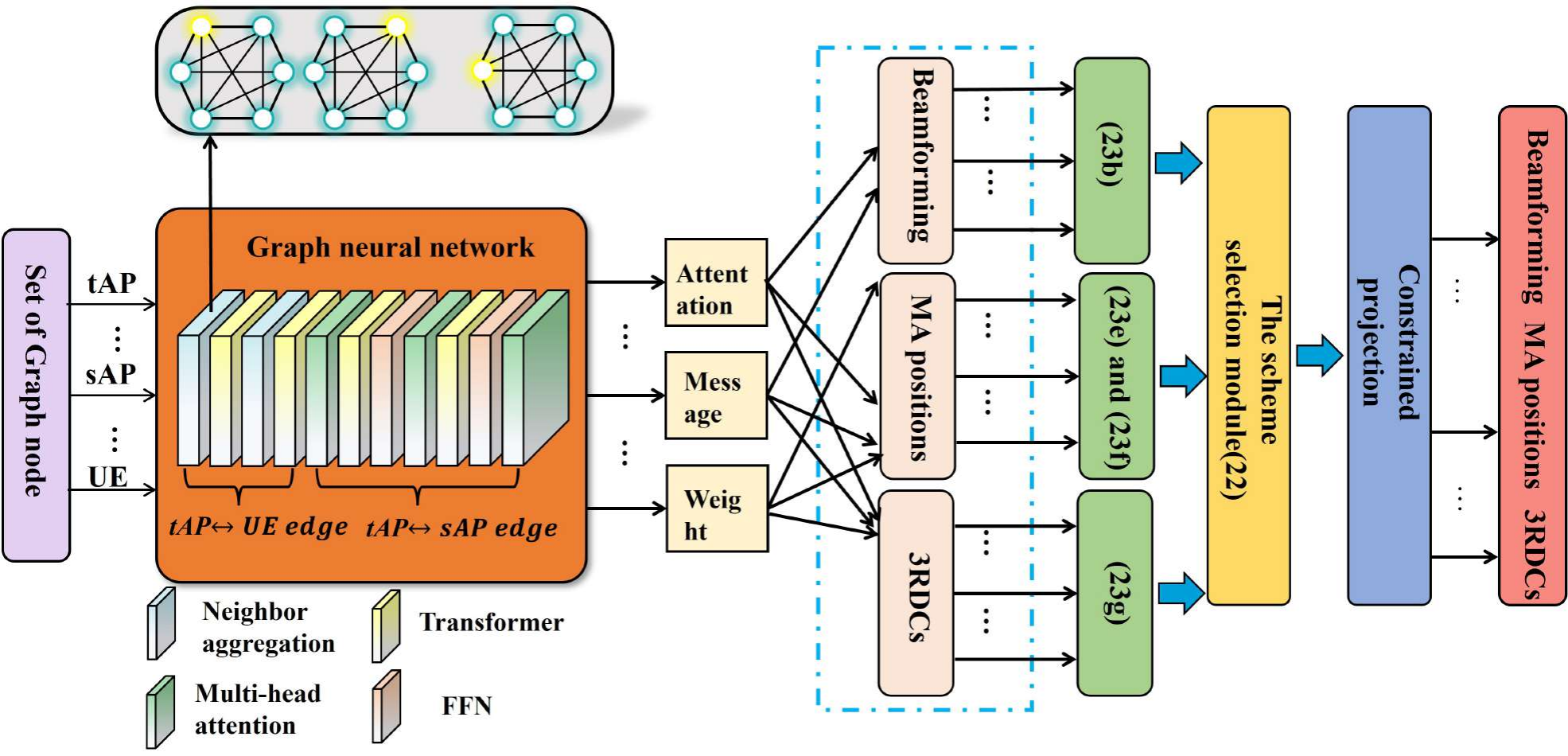}
  \captionsetup{justification=centering}
  \caption{The illustration of the architecture of the SACGNN.}\vspace{-10pt}
\label{FIGURETS__2}
\end{figure}

Training the SACGNN is performed by minimizing a composite loss function that penalizes violations of constraints
and promotes high sum-rate under distortion-aware conditions,
where $\mathcal{E}$ denotes the uncertainty set for distortion and channel
errors. $\mathbf{p}_{a}^{prev}$ and $\mathbf{p}_{b}^{prev}$ denote previous MA positions. The
architecture of the SACGNN is illustrated in Fig. 2.
 Fig.~\ref{FIGURETS__2}.

\subsection{Computational Complexity}
Let $A$ and $B$ denote the numbers of tAPs and sAPs, each equipped with $M_t$ and $M_s$ movable antennas (MAs), respectively, and $K$ denote the number of UEs. The heterogeneous graph then contains $V=AM_t+BM_s+K$ nodes and $E=E_{\mathrm{comm}}+E_{\mathrm{sens}}$ edges, where $E_{\mathrm{comm}}=\mathcal{O}(AM_tK)$ and $E_{\mathrm{sens}}=\mathcal{O}(AM_tBM_s)$. For an $L$-layer SACGNN with $H$ attention heads, model dimension $d_{\mathrm{model}}$, feed-forward width $d_{\mathrm{ff}}$, and edge-embedding dimension $d_e$, the per-layer complexity is $\mathcal{O}(V d_{\mathrm{model}} d_{\mathrm{ff}} + E H d_{\mathrm{head}}(1+d_e))$. Evaluating CRLB-based robust constraints requires $\mathcal{O}(S E_{\mathrm{sens}} d_\theta^3)$ under $S$ uncertainty realizations and per-target parameter dimension $d_\theta$. Therefore, the overall forward complexity is
\[
\mathcal{O}\!\Big(L(V d_{\mathrm{model}} d_{\mathrm{ff}} + E H d_{\mathrm{head}}(1+d_e)) + S E_{\mathrm{sens}} d_\theta^3\Big),
\]
which scales linearly with the number of graph edges.

\section{Numerical Results}\label{V}
\begin{figure}[!t]
  \centering
  \includegraphics[scale=0.28]{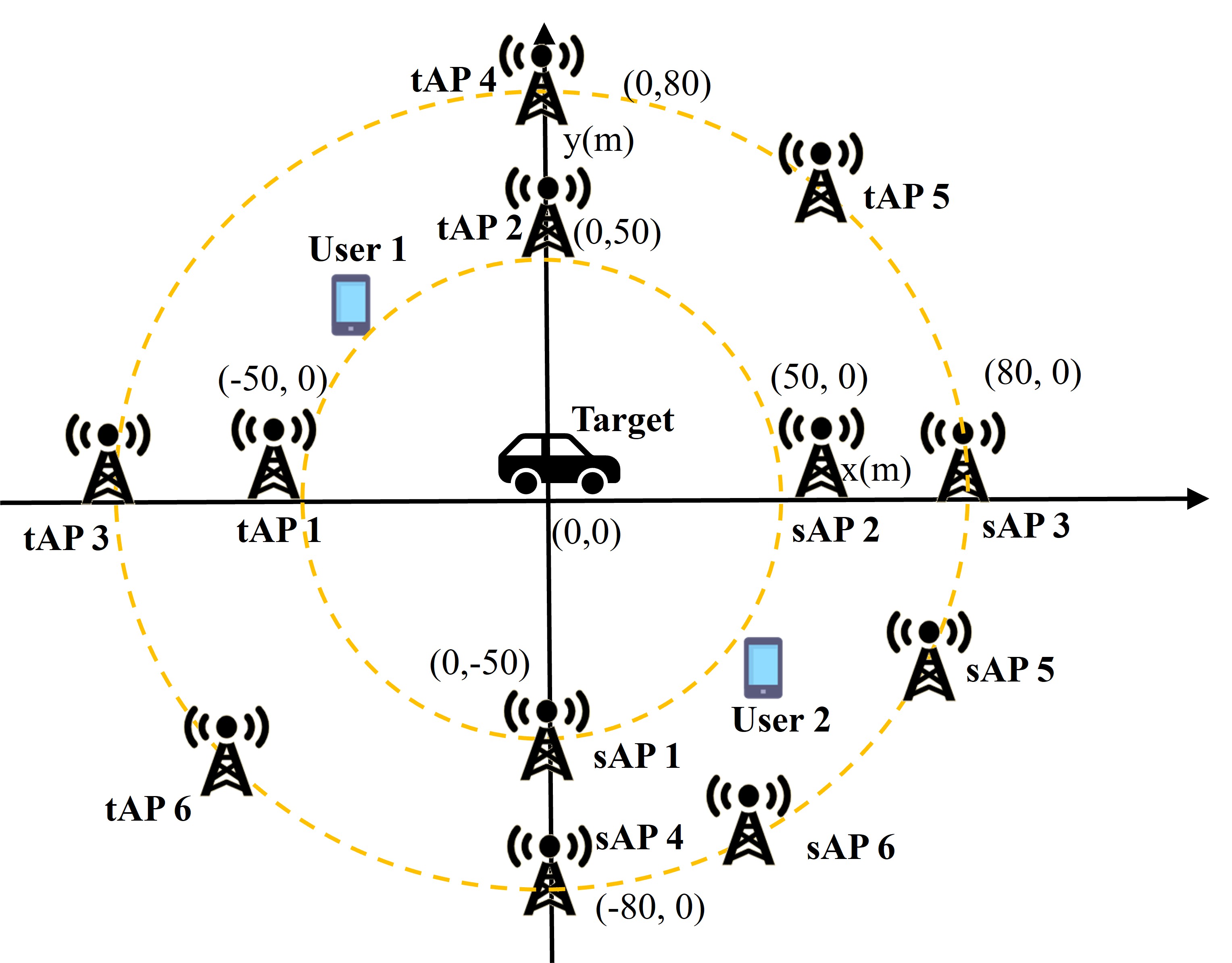}
  \captionsetup{justification=centering}
  \caption{Simulation setup of the MA-aided CF-ISAC system.}\vspace{-10pt}
\label{FIGURETS2}
\end{figure}
\subsection{Simulation Setup}
The sensing target is located at the origin, while $A=6$ tAPs, $B{=}6$ sAPs, and $U{=}2$ users are uniformly placed along a circular ring of radius $50$\,m (see Fig.~\ref{FIGURETS2}).  Each AP is equipped with $N{=}16$ transmit MAs and $M{=}4$ receive MAs, with minimum spacing $D_0{=}\lambda/2$. The transmit and receive MA ranges are set to $[-2\lambda, 2\lambda]$. The channel path loss follows $PL(d) = PL_0 (d/d_0)^{-\Omega}$, with $PL_0=-30$\,dB, $d_0=1$\,m, and path loss exponents $\Omega{=}2.8$ (AP-user) and $\Omega{=}2.2$ (AP-target). Each AP-UE link includes $L_{b,u}{=}3$ multipaths, and noise power is fixed at $-120$\,dBm. The target's RCS is $\alpha=3$, and required sensing accuracy is $\gamma_b=0.05$. The GNN at each AP uses $L{=}4$ graph convolution layers. All MLPs in the architecture share the same structure: $1600{\times}800$ fully connected layers with ReLU activation\cite{10632064}. The model is trained using the Adam optimizer with an initial learning rate of $0.01$, decayed by $0.995$ every 100 steps\cite{10632064}. Each epoch processes $50000$ samples with batch size $500$, totaling $100$ steps. Training stops after $5000$ epochs or if validation loss stagnates for $10$ epochs. The SOCP problem $\mathcal{P}$ is solved via MOSEK with bisection search, terminating when the loss change is below $0.01$. All algorithms are implemented in Python 3 using PyTorch Geometric, and simulations are run on an Intel Core i7 CPU with 16GB RAM under Windows 10. To benchmark the proposed SACGNN-robust method, we compare the following schemes:
\begin{itemize}
    \item \textbf{Baseline 1:} DRL with 3RDCs error-aware optimization\cite{10632064}.
    \item \textbf{Baseline 2:} DRL without 3RDCs error robustness\cite{10632064}.
    \item \textbf{Baseline 3:} SACGNN without 3RDCs error robustness.
    \item \textbf{FPA Algorithm:} SCA-based beamforming under fixed-position antenna architecture\cite{xiu2025movable}.
\end{itemize}

\begin{figure}[!t]
\centering
\begin{minipage}[t]{0.48\textwidth}
\centering
\includegraphics[width=0.7\textwidth, height=0.48\textwidth]{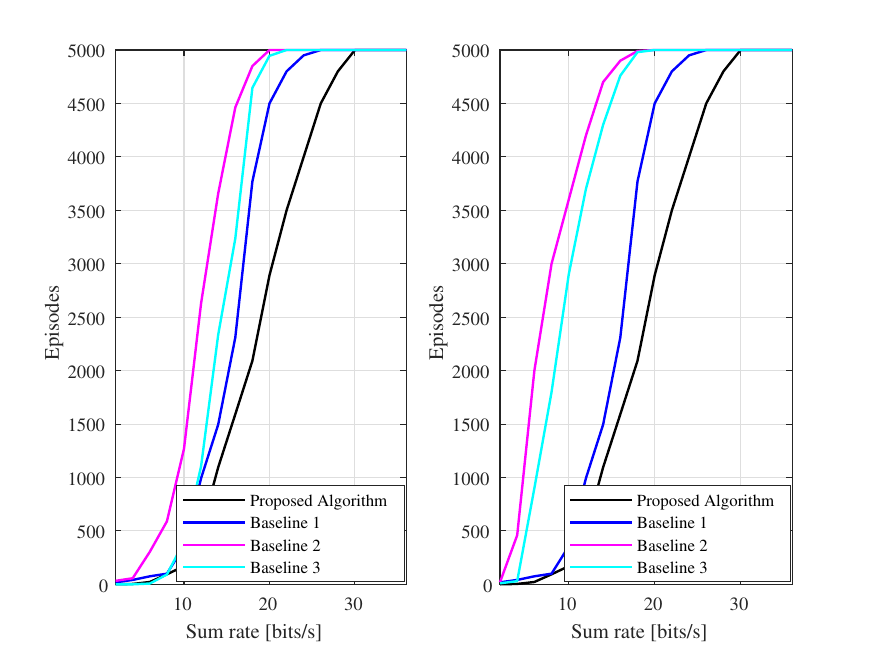}
\put(-130,-5){\small\textbf{(a)}}
\put(-50,-5){\small\textbf{(b)}}
\caption{ (a) Episodes versus sum rate without PA distortion. (b) Episodes versus sum rate with PA distortion.}\vspace{-10pt}
\label{FIGURETS4}
\end{minipage}
\end{figure}
As shown in Fig.~\ref{FIGURETS4}(a), for the PA distortion-free case, all methods initially show similar performance in the high-rate regime. However, as the sum rate decreases, the number of episodes of the baselines rapidly increases, particularly Baseline 2 and Baseline 3. In contrast, the proposed robust SACGNN and DRL maintains a significantly slower descent, indicating higher stability across varying conditions.  As shown in Fig.~\ref{FIGURETS4}(b), for the PA-distorted case,  the baselines—especially baselines 2 and 3—exhibit a steep drop in the number of episodes with high sum rate, demonstrating a clear lack of robustness to hardware-induced nonlinearities. Interestingly, Baseline 1 and proposed algorithm perform noticeably better than Baselines 2 and 3, confirming that the worst-case robust scheme. In addition, in Fig.~\ref{FIGURETS4}(a) and Fig.~\ref{FIGURETS4}(b), even small modeling mismatches can accumulate across distributed antennas and degrade sum-rate significantly. These results jointly show that the importance of robust graph-based policy learning in cell-free ISAC systems. Unlike conventional DRL methods, which often overfit to nominal conditions, the proposed robust SACGNN ensures reliable performance even in the presence of compound uncertainties arising from 3RDCs errors and hardware nonidealities.

\begin{figure}[!t]
\centering
\begin{minipage}[t]{0.48\textwidth}
\centering
\includegraphics[width=0.65\textwidth, height=0.45\textwidth]{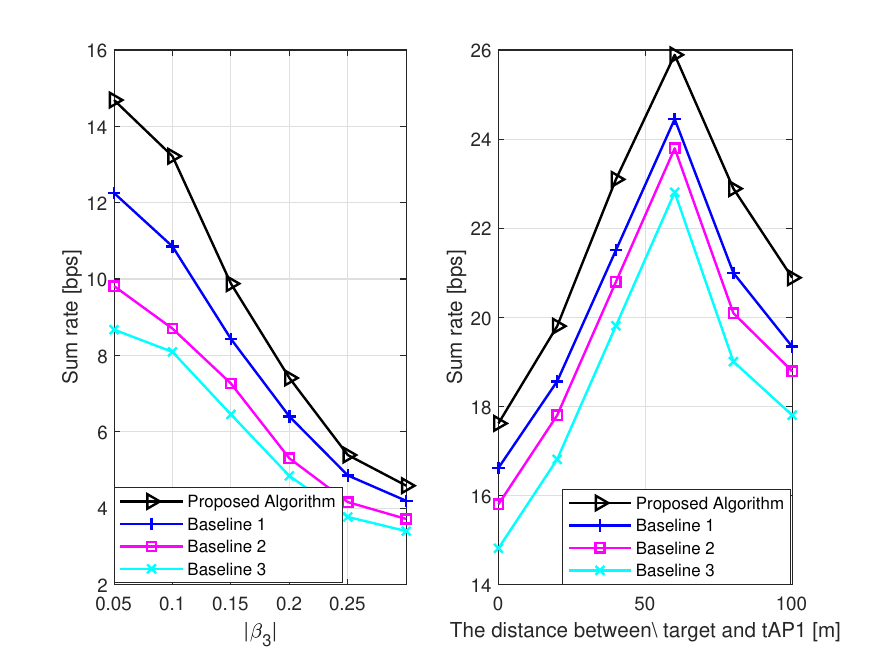}
\put(-120,-5){\small\textbf{(a)}}
\put(-50,-5){\small\textbf{(b)}}
\caption{(a) Sum rate versus the bound of $\beta_{3,a}$. (b) Sum rate versus the distance.}\vspace{-10pt}
\label{FIGURETS3}
\end{minipage}
\end{figure}

Fig.~\ref{FIGURETS3} depicts the robustness and adaptability of the proposed worst-case SACGNN optimization framework under both channel estimation uncertainty and hardware nonidealities. In particular, Fig.~\ref{FIGURETS3}(a) shows the sum rate performance versus the increasing normalized 3RDCs estimation error bound $|\beta_3|$, while Fig.~\ref{FIGURETS3}(b) illustrates the impact of varying target-to-access-point distance on the system sum rate.  In Fig.~\ref{FIGURETS3}(a), the proposed robust SACGNN clearly outperforms all baselines as the error bound $|\beta_3|$ increases. This trend reveals that non-robust methods fail to adapt to deteriorating 3RDCs, leading to a sharp decrease in achievable sum rates. The performance gap is especially notable when $|\beta_3| > 0.15$, where baseline methods suffer significant degradation due to overconfidence in nominal channel estimates. In Fig.~\ref{FIGURETS3}(b), the proposed algorithm consistently delivers superior performance across the entire distance range, even as the target moves farther from its nearest tAP.  As distance increases, the received signal strength weakens and the impact of PA distortion becomes more pronounced. For non-robust baselines, this dual degradation (weaker SNR + distorted amplification) leads to severe performance drops. Another key observation is that the performance curves of all baseline methods peak and then decline, reflecting their sensitivity to the compounded effects of large-scale fading and hardware nonidealities.

\begin{figure}[!t]
\centering
\begin{minipage}[t]{0.48\textwidth}
\centering
\includegraphics[width=0.65\textwidth, height=0.45\textwidth]{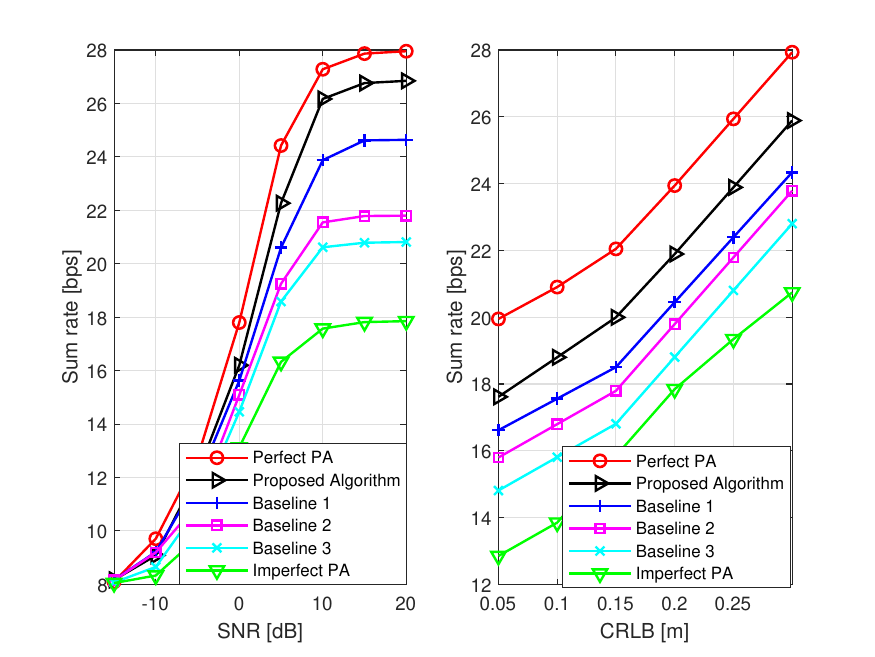}
\put(-120,-5){\small\textbf{(a)}}
\put(-50,-5){\small\textbf{(b)}}
\caption{(a) Sum rate versus SINDR. (b) Sum rate versus CRLB constraint.}\vspace{-10pt}
\label{FIGURETS6}
\end{minipage}
\end{figure}
Fig.~\ref{FIGURETS6} compares the sum rate performance of the proposed robust SACGNN framework against three baseline schemes, under both ideal and non-ideal PA (power amplifier) conditions. Fig.\ref{FIGURETS6}(a) shows the system’s performance versus SNR, while Fig.\ref{FIGURETS6}(b) illustrates the sensitivity to CRLB, which reflects sensing accuracy. In Fig.~\ref{FIGURETS6}(a), the Perfect PA scheme provides an upper bound where no hardware distortion is present. The proposed method closely approaches this bound across the entire SINDR range. As SNR increases, the baseline methods initially improve but gradually saturate or decline in slope. The Imperfect PA curve reflects the system performance under severe PA distortion without any compensation; its sharp deviation from the upper bound confirms the critical impact of hardware non-idealities. Notably, the proposed robust SACGNN maintains high-rate performance even under imperfect PA, due to its worst-case modeling of both estimation errors and amplifier distortion. In Fig.~\ref{FIGURETS6}(b), a similar performance trend emerges. While all methods exhibit increasing sum rate as CRLB increases, the proposed method consistently remains closest to the upper bound. This behavior may appear counterintuitive, as higher CRLB implies poorer sensing accuracy. However, a larger CRLB relaxes the sensing constraint, effectively loosening the joint ISAC optimization burden. As a result, more degrees of freedom—such as transmit power, beamforming, and phase control—can be allocated to communication, thereby enabling higher achievable rates. Baselines that ignore 3RDCs uncertainty (Baselines 2 and 3) degrade more rapidly, confirming that beamforming and decoding policies trained on nominal or unmodeled conditions lack generalizability.

\begin{figure}[!t]
  \centering
  \includegraphics[scale=0.35]{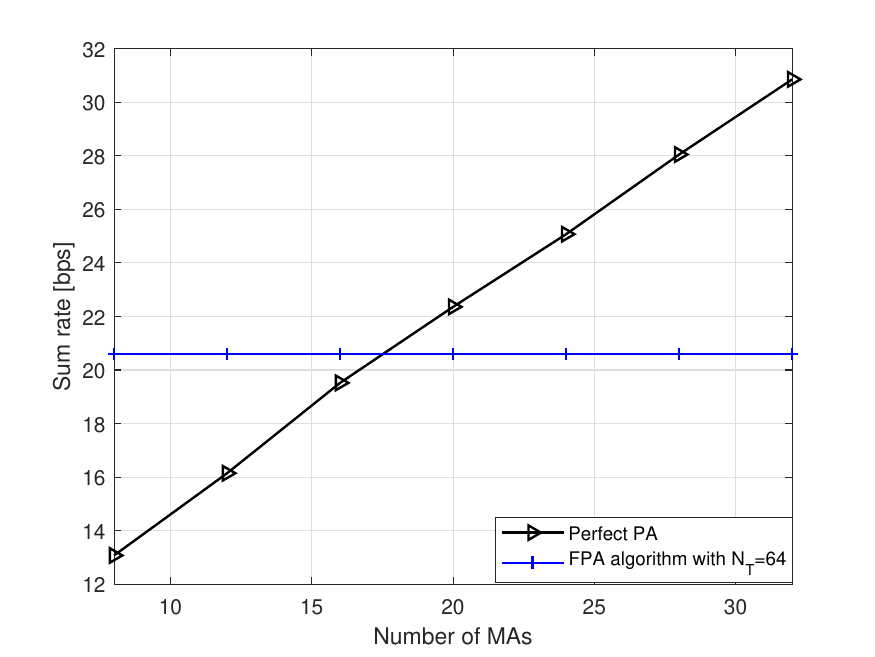}
  \captionsetup{justification=centering}
  \caption{Sum rate versus the number of MAs.}\vspace{-10pt}
\label{FIGURETS_2}
\end{figure}
Fig.~\ref{FIGURETS_2} compares the sum rate of the proposed robust SACGNN algorithm under perfect PA conditions with that of an FPA benchmark using SCA-based beamforming and $N_T=64$. As the number of MAs at the tAP increases from $8$ to $32$, the proposed algorithm achieves a steadily increasing sum rate, benefiting from additional spatial degrees of freedom and adaptive antenna positioning under joint communication and sensing constraints. By contrast, the FPA benchmark remains constant because its antenna configuration and element count are fixed. These results highlight the benefits of jointly optimizing antenna positions and beamforming in MA-aided CF-ISAC systems.

\section{Conclusion}\label{VI}
In this paper, we have developed a distributed distortionaware robust optimization framework for an MA-aided CFISAC system in the presence of nonlinear hardware impairments. Specifically, we have modeled the 3RDCs of power
amplifiers using norm-bounded uncertainty sets and addressed
their impact on both communication and sensing performance.
To mitigate these distortions, we have exploited the spatial
degrees of freedom offered by MAs, enabling the joint optimization of beamforming and antenna positioning. We have
proposed the SACGNN scheme to achieve robust and scalable
optimization in a decentralized fashion, accommodating the
distributed architecture of CF systems. Extensive simulations
have demonstrated that our proposed robust SACGNN algorithm significantly enhances both sensing accuracy and
communication QoS compared to non-robust baselines and
convex-approximation-based methods. The results have further
shown that adaptive MA positioning is particularly effective
in counteracting distortion effects, especially in high-mobility
or highly uncertain environments. Our framework provides
valuable insights into the design of intelligent, resilient, and
energy-efficient ISAC systems for beyond-5G and the 6G
wireless networks.

\begin{appendices}
\section{The proof of (\ref{pro14})}\label{APP1}
According to\cite{xiu2025movable}, $(n_{1},n_{2})$-th $(1\leq n_{1},n_{2}\leq 2)$ the FIM $\mathbf{J}_{b}(\mathbf{e})$ is denoted as
\begin{small}
\begin{align}
&F_{n_{1},n_{2}}^{b}=-\mathbb{E}_{\mathbf{y}_{b}|\mathbf{e}}\left(\frac{\partial^{2}\ln f(\mathbf{y}_{b}|\mathbf{e})}{\partial e_{n_{1}}\partial e_{n_{2}}}\right)=2\mathrm{Re}\nonumber\\
&\left\{\frac{\partial\sum\limits_{a=1}^{A}\mathbf{H}_{a,b}(\boldsymbol{\Xi}_{a}\mathbf{x}_{a}+\mathbf{d}_{a})}{\partial e_{n_{1}}}(\boldsymbol{\Upsilon}_{b})^{-1}\frac{\partial\sum\limits_{a=1}^{A}\mathbf{H}_{a,b}(\boldsymbol{\Xi}_{a}\mathbf{x}_{a}+\mathbf{d}_{a})}{\partial e_{n_{2}}}\right\}+\nonumber\\
&\mathrm{Tr}\left(\boldsymbol{\Upsilon}_{b}^{-1}\frac{\partial\sum\limits_{a=1}^{A}\mathbf{H}_{a,b}(\boldsymbol{\Xi}_{a}\mathbf{x}_{a}+\mathbf{d}_{a})}{\partial e_{n_{1}}}(\boldsymbol{\Upsilon}_{b})^{-1}\right.\left.\frac{\partial\sum\limits_{a=1}^{A}\mathbf{H}_{a,b}(\boldsymbol{\Xi}_{a}\mathbf{x}_{a}+\mathbf{d}_{a})}{\partial e_{n_{2}}}\right)\nonumber\\
&=\frac{2}{\sigma_{b}^{2}}\mathrm{Re}\left\{\frac{\partial\sum\limits_{a=1}^{A}\mathbf{H}_{a,b}(\boldsymbol{\Xi}_{a}\mathbf{x}_{a}+\mathbf{d}_{a})}{\partial e_{n_{1}}}\frac{\partial\sum\limits_{a=1}^{A}\mathbf{H}_{a,b}(\boldsymbol{\Xi}_{a}\mathbf{x}_{a}+\mathbf{d}_{a})}{\partial e_{n_{2}}}\right\},\label{proA1}%
\end{align}    
\end{small}%
where the first-order derivative of $\sum_{a=1}^{A}\mathbf{H}_{a,b}(\boldsymbol{\Xi}_{a}\mathbf{x}_{a}+\mathbf{d}_{a})$ and $\sum_{a=1}^{A}\mathbf{H}_{a,b}(\boldsymbol{\Xi}_{a}\mathbf{x}_{a}+\mathbf{d}_{a})$ withe respect to $e_{1}$ and $e_{2}$ are given by, respectively
\begin{align}
&\frac{\partial\sum_{a=1}^{A}\mathbf{H}_{a,b}(\boldsymbol{\Xi}_{a}\mathbf{x}_{a}+\mathbf{d}_{a})}{\partial e_{1}}=\sum\nolimits_{b=1}^{B}\dot{\mathbf{H}}_{a,b}(\boldsymbol{\Xi}_{a}\mathbf{x}_{a}+\mathbf{d}_{a}),\nonumber\\
&\frac{\partial\sum_{a=1}^{A}\mathbf{H}_{a,b}(\boldsymbol{\Xi}_{a}\mathbf{x}_{a}+\mathbf{d}_{a})}{\partial e_{2}}=\sum\nolimits_{b=1}^{B}\ddot{\mathbf{H}}_{a,b}(\boldsymbol{\Xi}_{a}\mathbf{x}_{a}+\mathbf{d}_{a}),\label{proA3}
\end{align}
where $\dot{\mathbf{H}}_{a,b}$ and $\ddot{\mathbf{H}}_{a,b}$ are denoted as
\begin{align}
&\dot{\mathbf{H}}_{a,b}=\zeta_{a,b}\dot{\mathbf{a}}_{r}(\varphi_{b})\mathbf{a}_{t}^{H}(\vartheta_{a})+\zeta_{a,b}\mathbf{a}_{r}(\varphi_{b})\dot{\mathbf{a}}_{t}^{H}(\vartheta_{a}),\nonumber\\
&\ddot{\mathbf{H}}_{a,b}=\zeta_{a,b}\ddot{\mathbf{a}}_{r}(\varphi_{b})\mathbf{a}_{t}^{H}(\vartheta_{a})+\zeta_{a,b}\mathbf{a}_{r}(\varphi_{b})\ddot{\mathbf{a}}_{t}^{H}(\vartheta_{a}).\label{proA4}
\end{align}
In (\ref{proA4}), $\dot{\mathbf{a}}_{t}(\vartheta_{a})$, $\dot{\mathbf{a}}_{r}(\varphi_{b})$, $\ddot{\mathbf{a}}_{t}(\vartheta_{a})$ and $\ddot{\mathbf{a}}_{r}(\varphi_{b})$ are denoted as
\begin{align}
&\dot{\mathbf{a}}_{t}^{H}(\phi_{a})=\dot{\phi}_{a}\tilde{\mathbf{a}}_{t}^{H}(\phi_{a})\odot\mathbf{a}_{t}^{H}(\phi_{a}),\dot{\mathbf{a}}_{r}(\varphi_{b})=\dot{\varphi}_{b}\tilde{\mathbf{a}}_{r}(\varphi_{b})\odot\mathbf{a}_{r}(\varphi_{b}),\nonumber\\
&\ddot{\mathbf{a}}_{t}^{H}(\phi_{a})=\ddot{\phi}_{a}\tilde{\mathbf{a}}_{t}^{H}(\phi_{a})\odot\mathbf{a}_{t}^{H}(\phi_{a}),\ddot{\mathbf{a}}_{r}(\varphi_{b})=\ddot{\varphi}_{b}\tilde{\mathbf{a}}_{r}(\varphi_{b})\odot\mathbf{a}_{r}(\varphi_{b}),\label{proA5}
\end{align} 
where
$\dot{\phi}_{a}=(d_{y}^{a}-d_{y})/((d_{x}^{a}-d_{x})^{2}+(d_{y}^{a}-d_{y})^{2})$,$\dot{\varphi}_{b}=(d^{b}_{y}-d_{y})/((d^{b}_{x}-d_{x})^{2}+(d^{b}_{y}-d_{y})^{2})$,$\ddot{\phi}_{a}=(d_{x}^{b}-d_{x})/((d_{x}^{b}-d_{x})^{2}+(d_{y}^{b}-d_{y})^{2})$,$\ddot{\varphi}_{b}=(d^{b}_{y}-d_{y})/((d^{b}_{x}-d_{x})^{2}+(d^{b}_{y}-d_{y})^{2})$. 
According to (\ref{proA5}), we have
\begin{align}
&F_{n_{1},n_{2}}^{b}=\frac{2}{\sigma_{b}^{2}}\nonumber\\
&\mathrm{Re}\left\{
\sum_{a=1}^{A}\sum_{\bar{a}=1}^{A}\mathrm{Tr}\left(\mathbf{G}_{b,a,\bar{a}}^{n_{1},n_{2}}(\boldsymbol{\Xi}_{a}\mathbf{x}_{a}+\mathbf{d}_{a})(\mathbf{G}_{\bar{a}}\mathbf{x}_{\bar{a}}+\mathbf{d}_{\bar{a}})^{H}\right)\right\},\label{proA7}
\end{align}
where 
\begin{align}
\mathbf{G}_{b,a,\bar{a}}^{n_{1},n_{2}}=\left\{
\begin{matrix}
\dot{\mathbf{H}}_{a,b}^{H}\dot{\mathbf{H}}_{\bar{a},b}&n_{1}=1,n_{2}=1,\\
\dot{\mathbf{H}}_{a,b}^{H}\ddot{\mathbf{H}}_{\bar{a},b}&n_{1}=1,n_{2}=2,\\
\ddot{\mathbf{H}}_{a,b}^{H}\dot{\mathbf{H}}_{\bar{a},b}&n_{1}=2,n_{2}=1,\\
\ddot{\mathbf{H}}_{a,b}^{H}\ddot{\mathbf{H}}_{\bar{a},b}&n_{1}=2,n_{2}=2.
\end{matrix}\right.
\end{align}
Then, we expand the expression based on the Bussgang decomposition, (\ref{proA7}), and (\ref{proA8}) is given at the top of this page.
\begin{figure*}
\begin{small}
\begin{align}
&\mathbb{E}\left\{F_{n_{1},n_{2}}^{b}\right\}=\sum\nolimits_{a=1}^{A}\sum\nolimits_{\bar{a}=1}^{A}\mathrm{Re}\left\{\beta_{1}^{2}\underbrace{\mathrm{Tr}\left(\mathbf{G}_{b,a,\bar{a}}^{n_{1},n_{2}}\mathbf{W}_{a}\mathbf{W}_{\bar{a}}^{H}\right)}_{\dot{\Upsilon}_{b,a,\bar{a}}^{n_{1},n_{2}}}+2(\bar{\beta}_{3,a}+\Delta\beta_{3,a})\beta_{1}\underbrace{\mathrm{Tr}\left(\mathbf{G}_{b,a,\bar{a}}^{n_{1},n_{2}}\mathrm{diag}\{\mathbf{W}_{a}\mathbf{W}_{a}^{H}\}\mathbf{W}_{a}\mathbf{W}_{\bar{a}}^{H}\right)}_{\ddot{\Upsilon}_{b,a,\bar{a}}^{n_{1},n_{2}}}+2\beta_{3,\bar{a}}\right.\nonumber\\
&\left.\beta_{1}\underbrace{\mathrm{Tr}\left(\mathbf{G}_{b,a,\bar{a}}^{j_{1},j_{2}}\mathbf{W}_{a}\mathbf{W}_{\bar{a}}^{H}(\mathrm{diag}\{\mathbf{W}_{\bar{a}}\mathbf{W}_{\bar{a}}^{H}\})^{H}\right)}_{\bar{\Upsilon}_{b,a,\bar{a}}^{n_{1},n_{2}}}+4(\bar{\beta}_{3,a}+\Delta\beta_{3,a})\beta_{3,\bar{a}}\underbrace{\mathrm{Tr}\left(\mathrm{diag}\{\mathbf{W}_{\bar{a}}\mathbf{W}_{\bar{a}}\}\mathbf{W}_{a}\mathbf{W}_{\bar{a}}^{H}(\mathrm{diag}\{\mathbf{W}_{\bar{a}}\mathbf{W}_{\bar{a}}^{H}\})^{H}\right)}_{\tilde{\Upsilon}_{b,a,\bar{a}}^{n_{1},n_{2}}}\right\}\nonumber\\
&+\sum\nolimits_{a=1}^{A}2|(\bar{\beta}_{3,a}+\Delta\beta_{3,a})|^{2}\underbrace{\mathrm{Tr}(\mathbf{W}_{a}\mathbf{W}_{a}^{H}\odot|\mathbf{W}_{a}\mathbf{W}_{a}^{H}|^{2})}_{\Upsilon_{a}}=\sum_{a=1}^{A}\sum_{\bar{a}=1}^{A}\mathrm{Re}\{\beta_{1}^{2}\dot{\Upsilon}_{b,a,\bar{a}}^{n_{1},n_{2}}+2(\bar{\beta}_{3,a}+\Delta\beta_{3,a})\beta_{1}\ddot{\Upsilon}_{b,a,\bar{a}}^{n_{1},n_{2}}+2\beta_{3,\bar{a}}\beta_{1}\bar{\Upsilon}_{b,a,\bar{a}}^{n_{1},n_{2}}\nonumber\\
&+4(\bar{\beta}_{3,a}+\Delta\beta_{3,a})\beta_{3,\bar{a}}\tilde{\Upsilon}_{b,a,\bar{a}}^{n_{1},n_{2}}\}+\sum_{a=1}^{A}2|(\bar{\beta}_{3,a}+\Delta\beta_{3,a})|^{2}\Upsilon_{a}.\label{proA8}
\end{align}    
\end{small}
\hrulefill
\end{figure*}
According to the triangle inequality, we have
\begin{align}
&\mathbb{E}\left\{F_{n_{1},n_{2}}^{b}\right\}\leq c+|(\bar{\beta}_{3,a}+\Delta\beta_{3,a})||\ddot{\Upsilon}_{b,a,\bar{a}}^{n_{1},n_{2}}|+|(\bar{\beta}_{3,a}+\Delta\beta_{3,a})|\nonumber\\
&|\bar{\Upsilon}_{b,a,\bar{a}}^{n_{1},n_{2}}|+|(\bar{\beta}_{3,a}+\Delta\beta_{3,a})||\beta_{3,\bar{a}}||\tilde{\Upsilon}_{b,a,\bar{a}}^{n_{1},n_{2}}|\nonumber\\
&+\sum\nolimits_{a=1}^{A}2|(\bar{\beta}_{3,a}+\Delta\beta_{3,a})|^{2}\Upsilon_{a}.\label{proA9}    
\end{align}
and
\begin{align}
&\mathbb{E}\left\{F_{n_{1},n_{2}}^{b}\right\}\leq c+(\epsilon-\bar{\beta}_{3,a})|\ddot{\Upsilon}_{b,a,\bar{a}}^{n_{1},n_{2}}|+(\epsilon-\bar{\beta}_{3,a})|\bar{\Upsilon}_{b,a,\bar{a}}^{n_{1},n_{2}}|\nonumber\\
&+(\epsilon^{2}-\bar{\beta}_{3,a}^{2})|\tilde{\Upsilon}_{b,a,\bar{a}}^{n_{1},n_{2}}|+\sum\nolimits_{a=1}^{A}2(\epsilon^{2}-\bar{\beta}_{3,a}^{2})\Upsilon_{a}.\label{proA10}  
\end{align}
To simplify the notation, we define
\begin{align}
&\Xi=\bar{\beta}_{3,a}|\ddot{\Upsilon}_{b,a,\bar{a}}^{n_{1},n_{2}}|+\bar{\beta}_{3,a}|\bar{\Upsilon}_{b,a,\bar{a}}^{n_{1},n_{2}}|+\bar{\beta}_{3,a}^{2}|\tilde{\Upsilon}_{b,a,\bar{a}}^{n_{1},n_{2}}|\nonumber\\
&+\sum\nolimits_{a=1}^{A}2\bar{\beta}_{3,a}^{2}\Upsilon_{a}.
\end{align}
Similarly, the lower bound of $\mathbb{E}\left\{F_{n_{1},n_{2}}^{b}\right\}$ is given by
\begin{align}
&\mathbb{E}\left\{F_{n_{1},n_{2}}^{b}\right\}\geq c-\epsilon|\ddot{\Upsilon}_{b,a,\bar{a}}^{n_{1},n_{2}}|-\epsilon|\bar{\Upsilon}_{b,a,\bar{a}}^{n_{1},n_{2}}|-\epsilon^{2}|\tilde{\Upsilon}_{b,a,\bar{a}}^{n_{1},n_{2}}|\nonumber\\
&+\sum\nolimits_{a=1}^{A}2\epsilon^{2}\Upsilon_{a}-F(\ddot{\Upsilon}_{b,a,\bar{a}}^{n_{1},n_{2}},\bar{\Upsilon}_{b,a,\bar{a}}^{n_{1},n_{2}},\tilde{\Upsilon}_{b,a,\bar{a}}^{n_{1},n_{2}},\Upsilon_{a}).\label{proA11}   
\end{align}
The upper bound of $\mathbb{E}\left\{F_{1,1}^{b}\right\}+\mathbb{E}\left\{F_{2,2}^{b}\right\}$ is given by
\begin{align}
&\mathbb{E}\left\{F_{1,1}^{b}\right\}+\mathbb{E}\left\{F_{2,2}^{b}\right\}\leq c+\epsilon\Theta+\epsilon^{2}\dot{\Theta}+2\Pi.\label{proA_11}   
\end{align}
Similarly, the lower bound of $\mathbb{E}\left\{F_{1,1}^{b}\right\}\mathbb{E}\left\{F_{2,2}^{b}\right\}$ and the upper bound of $\mathbb{E}\left\{F_{1,2}^{b}\right\}\mathbb{E}\left\{F_{2,1}^{b}\right\}$ are given in (\ref{proA12}) at the top of next page.
\begin{figure*}
\begin{small}
\begin{align}
&\mathbb{E}\left\{F_{1,1}^{b}\right\}\mathbb{E}\left\{F_{2,2}^{b}\right\}\geq (c+\Xi)^{2} 
- \epsilon (c+2\Pi)(\Theta_1 + \Theta_2) 
+ \epsilon^2\left[(c+2\Pi)(\dot{\Theta}_1 + \dot{\Theta}_2) + \Theta_1 \Theta_2\right]-\epsilon^3 \left(\Theta_1 \dot{\Theta}_2 + \Theta_2 \dot{\Theta}_1\right)
+ \epsilon^4 \dot{\Theta}_1 \dot{\Theta}_2,\nonumber\\
&\mathbb{E}\left\{F_{1,2}^{b}\right\}\mathbb{E}\left\{F_{2,1}^{b}\right\}\leq (c+\Xi)^{2} 
+ \epsilon (c+2\Pi)(\Theta_3 + \Theta_4)+ \epsilon^2 \left[(c+2\Pi)(\dot{\Theta}_3 + \dot{\Theta}_4) + \Theta_3 \Theta_4\right]+ \epsilon^3 \left(\Theta_3 \dot{\Theta}_4 + \Theta_4 \dot{\Theta}_3\right) 
+ \epsilon^4 \dot{\Theta}_3 \dot{\Theta}_4,\label{proA12}   
\end{align}      
\end{small} 
\hrulefill
\end{figure*}
$\Theta=(|\ddot{\Upsilon}_{b,a,\bar{a}}^{1,1}|+|\ddot{\Upsilon}_{b,a,\bar{a}}^{2,2}|+|\bar{\Upsilon}_{b,a,\bar{a}}^{1,1}|+|\bar{\Upsilon}_{b,a,\bar{a}}^{2,2}|)$, 
$\dot{\Theta}=(|\tilde{\Upsilon}_{b,a,\bar{a}}^{1,1}|+|\tilde{\Upsilon}_{b,a,\bar{a}}^{2,2}|+\sum_{a=1}^{A}4\Upsilon_{a})$
$\Theta_{1}(|\ddot{\Upsilon}_{b,a,\bar{a}}^{1,1}|+|\bar{\Upsilon}_{b,a,\bar{a}}^{1,1}|)$, $\dot{\Theta}_{1}=(\sum_{a=1}^{A}2\Upsilon_{a}-|\tilde{\Upsilon}_{b,a,\bar{a}}^{1,1}|)$, $\Theta_{2}=(|\ddot{\Upsilon}_{b,a,\bar{a}}^{2,2}|+|\bar{\Upsilon}_{b,a,\bar{a}}^{2,2}|)$, $\dot{\Theta}_{2}=(\sum_{a=1}^{A}2\Upsilon_{a}-|\tilde{\Upsilon}_{b,a,\bar{a}}^{2,2}|)$, $\Theta_{3}=(|\ddot{\Upsilon}_{b,a,\bar{a}}^{1,2}|+|\bar{\Upsilon}_{b,a,\bar{a}}^{1,2}|)$, $\Theta_{4}=(|\ddot{\Upsilon}_{b,a,\bar{a}}^{2,1}|+|\bar{\Upsilon}_{b,a,\bar{a}}^{2,1}|)$, $\dot{\Theta}_{3}=(\sum_{a=1}^{A}2\Upsilon_{a}+|\tilde{\Upsilon}_{b,a,\bar{a}}^{1,2}|)$, $\dot{\Theta}_{4}=(\sum_{a=1}^{A}2\Upsilon_{a}+|\tilde{\Upsilon}_{b,a,\bar{a}}^{2,1}|)$.
Therefore, for improved robustness, we scale the original constraint in (\ref{pro21c}), leading to a stricter but more conservative formulation, and it is expressed as
\begin{align}
& c + \epsilon \Theta + \epsilon^2 \dot{\Theta}+2\Pi\leq g(\epsilon),
\end{align}
where $g(\epsilon)$ is expressed as
\begin{align}
&g(\epsilon,\epsilon_{0})=\gamma_b \bigg[ 
- \epsilon (c+2\Pi)(\Theta_1 + \Theta_2) 
+ \epsilon^2 \left( (c+2\Pi)(\dot{\Theta}_1\right.\nonumber\\
&\left.+ \dot{\Theta}_2) + \Theta_1 \Theta_2 \right)-\left(\epsilon^3\right)(\Theta_1 \dot{\Theta}_2 + \Theta_2 \dot{\Theta}_1)+ \left(\epsilon^4 \right)\dot{\Theta}_1 \dot{\Theta}_2\nonumber\\ 
&-\epsilon (c+2\Pi)(\Theta_3+\Theta_4)-\epsilon^2 \left( (c+2\Pi)(\dot{\Theta}_3 + \dot{\Theta}_4) + \Theta_3 \Theta_4 \right)\nonumber\\
&- \left(\epsilon^3 \right)(\Theta_3 \dot{\Theta}_4 + \Theta_4 \dot{\Theta}_3)-\left(\epsilon^4\right) \dot{\Theta}_3 \dot{\Theta}_4 
\bigg].
\end{align}
Since $g(\epsilon)$ is non-convex, we use SCA to obtain the first-order Taylor series expansion of $g(\epsilon)$ at $\epsilon_{0}$, and  it is denoted as $g(\epsilon,\epsilon_{0})$, and it is given in (\ref{pro32}).

\section{\texorpdfstring{Proof of Robust Scaling of $\delta$}{Proof of Robust Scaling}}
To enhance the robustness of the rate expression under 3RDCs, we scale the communication-related objective based on bounded distortion effects. The $\delta$ by nonlinear distortions is given by
\begin{align}
&\mathrm{Re}\{\zeta_{k}^{*}\sum\nolimits_{a=1}^{A}\mathbf{h}_{b,k}^{H}\boldsymbol{\Xi}_{b}\mathbf{w}_{b,k}\}=\mathrm{Re}\{\zeta_{k}^{*}\sum\nolimits_{a=1}^{A}\beta_{1}\mathbf{h}_{b,k}^{H}\mathbf{w}_{b,k}\}+\nonumber\\
&\mathrm{Re}\{\zeta_{k}^{*}\sum\nolimits_{a=1}^{A}2(\bar{\beta}_{3,a}+\Delta\beta_{3,a})\mathbf{h}_{a,k}^{H}\mathrm{diag}\{\mathbf{W}_{a}\mathbf{W}_{a}\}\mathbf{w}_{a,k}\}.\label{proB1}
\end{align}
Using the triangle inequality, we conservatively bound the second term caused by nonlinear distortion as
\begin{align}
&\mathrm{Re}\{\zeta_{k}^{*}\sum\nolimits_{a=1}^{A}2(\bar{\beta}_{3,a}+\Delta\beta_{3,a})\mathbf{h}_{a,k}^{H}\mathrm{diag}\{\mathbf{W}_{a}\mathbf{W}_{a}\}\mathbf{w}_{a,k}\}\geq\nonumber\\
&-\sum\nolimits_{a=1}^{A}|(\bar{\beta}_{3,a}+\Delta\beta_{3,a})||\zeta_{k}^{*}2\mathbf{h}_{a,k}^{H}\mathrm{diag}\{\mathbf{W}_{a}\mathbf{W}_{a}\}\mathbf{w}_{a,k}|\nonumber\\
&\geq -\epsilon\sum\nolimits_{a=1}^{A}|\zeta_{k}^{*}2\mathbf{h}_{a,k}^{H}\mathrm{diag}\{\mathbf{W}_{a}\mathbf{W}_{a}\}\mathbf{w}_{a,k}|,\label{proB2} 
\end{align}
where $\epsilon$ is the upper bound on the 3RDCs, i.e., $|(\bar{\beta}_{3,a}+\Delta\beta_{3,a})| \leq \epsilon$.
Similarly, the interference term is upper bounded as
\begin{align}
&|\sum\nolimits_{b=1}^{B}\mathbf{h}_{b,k}^{H}\boldsymbol{\Xi}_{b}\mathbf{w}_{b,j}|^{2}\leq |\sum\nolimits_{b=1}^{B}\beta_{1}\mathbf{h}_{b,k}^{H}\mathbf{w}_{b,j}|^{2}+\nonumber\\
&\epsilon^{2}|\sum\nolimits_{b=1}^{B}2\mathbf{h}_{b,k}^{H}\mathrm{diag}\{\mathbf{W}_{a}\mathbf{W}_{a}^{H}\}\mathbf{w}_{b,j}|^{2}.\label{proB3}
\end{align}
The interference caused by the distorted signal component can also be bounded as
\begin{align}
&\sum\nolimits_{a=1}^{A}\mathbf{h}_{a,k}^{H}\boldsymbol{\Xi}_{b}\mathbf{h}_{a,k}\leq\sum\nolimits_{a=1}^{A}2\epsilon^{2}\mathbf{h}_{a,k}^{H}(\mathbf{W}_{a}\mathbf{W}_{a}^{H}\odot\nonumber\\
&|\mathbf{W}_{a}\mathbf{W}_{a}^{H}|^{2})\mathbf{h}_{a,k}.\label{proB4}
\end{align}
Combining the above conservative bounds, the lower bound of $\delta$ is scaled as follows:
\begin{align}
\delta\geq\mathcal{L}(\epsilon),\label{proB5}
\end{align}   
where $\mathcal{L}(\epsilon)$ is given in (\ref{pro32}).
\end{appendices}


\begin{thebibliography}{99}
\bibitem{9737357} F. Liu, Y. Cui, C. Masouros, J. Xu, T. X. Han, Y. C. Eldar, and S. Buzzi,
“Integrated sensing and communications: Toward dual-functional wire-
less networks for 6G and beyond,” IEEE J. Sel. Areas Commun., vol. 40,
no. 6, pp. 1728–1767, 2022.
\bibitem{9743355} E. Shi, J. Zhang, S. Chen, J. Zheng, Y. Zhang, D. W. Kwan Ng, and
B. Ai, “Wireless energy transfer in RIS-aided cell-free massive MIMO
systems: Opportunities and challenges,” IEEE Commun. Mag., vol. 60,
no. 3, pp. 26–32, 2022.
\bibitem{10286328} L. Zhu, W. Ma, and R. Zhang, “Movable antennas for wireless commu-
nication: Opportunities and challenges,” IEEE Commun. Mag., vol. 62,
no. 6, pp. 114–120, Oct. 2024.
\bibitem{9104787} T. Ackermann, J. Potschka, T. Maiwald, A. Hagelauer, G. Fischer, and
R. Weigel, “A robust digital predistortion algorithm for 5G MIMO:
Modeling a MIMO scenario with two nonlinear MIMO transmitters
including a cross-coupling effect,” IEEE Microwave Mag., vol. 21, no. 7,
pp. 54–62, 2020.
\bibitem{6954520} L. Guan and A. Zhu, “Green communications: Digital predistortion for
wideband RF power amplifiers,” IEEE Microwave Mag., vol. 15, no. 7,
pp. 84–99, 2014.
\bibitem{7432149} M. Abdelaziz, Z. Fu, L. Anttila, A. M. Wyglinski, and M. Valkama,
“Digital predistortion for mitigating spurious emissions in spectrally
agile radios,” IEEE Commun. Mag., vol. 54, no. 3, pp. 60–69, 2016.
\bibitem{10646258} A. R. Mohamed Ismail, M. Guenach, A. Sakhnini, A. Bourdoux, and
H. Steendam, “Robustness of ISAC waveforms to power amplifier
distortion,” in 2024 IEEE 4th International Symposium on Joint Com-
munications Sensing (JCS), 2024, pp. 1–6.
\bibitem{1255683} J. Vuolevi and T. Rahkonen, “Analysis of third-order intermodulation
distortion in common-emitter BJT and HBT amplifiers,” IEEE Trans.
Circuits Syst. II Express Briefs., vol. 50, no. 12, pp. 994–1001, 2003.
\bibitem{599544} M. Tummla, M. Donovan, B. Watkins, and R. North, “Volterra series
based modeling and compensation of nonlinearities in high power am-
plifiers,” in 1997 IEEE International Conference on Acoustics, Speech,
and Signal Processing, vol. 3, 1997, pp. 2417–2420 vol.3.
\bibitem{4717214} J. Mateu, C. Collado, N. Orloff, J. C. Booth, E. Rocas, A. Padilla,
and J. M. O’Callaghan, “Third-order intermodulation distortion and har-
monic generation in mismatched weakly nonlinear transmission lines,”
IEEE Trans. Microwave Theory Tech., vol. 57, no. 1, pp. 10–18, 2009.
\bibitem{1703853} D. Morgan, Z. Ma, J. Kim, M. Zierdt, and J. Pastalan, “A generalized
memory polynomial model for digital predistortion of rf power ampli-
fiers,” IEEE Trans. Signal Process., vol. 54, no. 10, pp. 3852–3860,
2006.
\bibitem{10906511} L. Zhu, W. Ma, W. Mei, Y. Zeng, Q. Wu, B. Ning, Z. Xiao, X. Shao, J. Zhang, and R. Zhang, “A tutorial on movable antennas for wireless networks,” IEEE Commun. Surv. Tutorials, vol. 28, pp. 3002--3054, 2026.
\bibitem{10146262} K.-K. Wong, K.-F. Tong, and C.-B. Chae, “Fluid antenna system—part
III: A new paradigm of distributed artificial scattering surfaces for
massive connectivity,” IEEE Commun. Lett., vol. 27, no. 8, pp. 1929–
1933, 2023.
\bibitem{10243545} W. Ma, L. Zhu, and R. Zhang, “MIMO capacity characterization for
movable antenna systems,” IEEE Trans. Wireless Commun., vol. 23,
no. 4, pp. 3392–3407, 2024.
\bibitem{10978811} S. Peng, C. Zhang, Y. Xu, Q. Wu, L. Zhu, X. Ou, and D. He, “Joint
antenna position and beamforming optimization with self-interference
mitigation in movable antenna aided ISAC system,” in 2025 IEEE
Wireless Communications and Networking Conference (WCNC), 2025,
pp. 1–6.
\bibitem{11018493} H. Wei, W. Wang, W. Ni, C. Zhang, and Y. Huang, “Movable-antenna
enabled cell-free networks,” IEEE Trans. Veh. Technol., pp. 1–6, 2025.
\bibitem{8606437} Y. Cai, Y. Xu, Q. Shi, B. Champagne, and L. Hanzo, “Robust joint
hybrid transceiver design for millimeter wave full-duplex MIMO relay
systems,” IEEE Trans. Wireless Commun., vol. 18, no. 2, pp. 1199–1215,
Jan. 2019.
\bibitem{9307295} O. T. Demir and E. Bjornson, “The bussgang decomposition of nonlinear
systems: Basic theory and MIMO extensions [lecture notes],” IEEE
Signal Process Mag., vol. 38, no. 1, pp. 131–136, 2021.
\bibitem{10869384} M. Liu, M. Li, R. Liu, and Q. Liu, “Distributed distortion-aware
beamforming designs for Cell-Free mMIMO systems,” IEEE Journal
of Selected Topics in Signal Processing, vol. 19, no. 2, pp. 381–397,
2025.
\bibitem{10032173} Y. Ju, H. Wang, Y. Chen, T.-X. Zheng, Q. Pei, J. Yuan, and N. Al-
Dhahir, “Deep reinforcement learning based joint beam allocation and
relay selection in mmWave vehicular networks,” IEEE Trans. Commun.,
vol. 71, no. 4, pp. 1997–2012, 2023.
\bibitem{7961152} K. Venugopal, A. Alkhateeb, N. Gonz´alez Prelcic, and R. W. Heath,
“Channel estimation for hybrid architecture-based wideband millimeter
wave systems,” IEEE J. Sel. Areas Commun., vol. 35, no. 9, pp. 1996–
2009, 2017.
\bibitem{ning2025movable} B. Ning, S. Yang, Y. Wu, P. Wang, W. Mei, C. Yuen, and E. Bjornson,
“Movable antenna-enhanced wireless communications: General archi-
tectures and implementation methods,” IEEE Wireless Commun., 2025.
\bibitem{9319879} W. Xia, Y. Li, and S. Dong, “Radar-based high-accuracy cardiac activity
sensing,” IEEE Transactions on Instrumentation and Measurement,
vol. 70, pp. 1–13, 2021.
\bibitem{10662951} S. Kang, K. Han, and S. Hong, “Sensing-aided distortion estimation for
ofdm radar with nonlinear transmitter,” IEEE Transactions on Radar
Systems, vol. 2, pp. 821–831, 2024.
\bibitem{9472958} Y. Xiu, J. Zhao, W. Sun, M. D. Renzo, G. Gui, Z. Zhang, and N. Wei,
“Reconfigurable intelligent surfaces aided mmWave NOMA: Joint power
allocation, phase shifts, and hybrid beamforming optimization,” IEEE
Trans. Wireless Commun., vol. 20, no. 12, pp. 8393–8409, Jul. 2021.
\bibitem{9348933} Y. Xiu, J. Zhao, W. Sun, and Z. Zhang, “Secrecy rate maximization
for reconfigurable intelligent surface aided millimeter wave system with
low-resolution DACs,” IEEE Commun. Lett., vol. 25, no. 7, pp. 2166–
2170, Feb. 2021.
\bibitem{10035459} W. Lyu, Y. Xiu, S. Yang, C. Yuen, and Z. Zhang, “Energy-efficient cell-
free network assisted by hybrid RISs,” IEEE Wireless Commun. Lett.,
vol. 12, no. 4, pp. 718–722, Feb. 2023.
\bibitem{10027476} Y. Huang, H. Fu, S. A. Vorobyov, and Z.-Q. Luo, “Robust adaptive
beamforming via Worst-Case SINR maximization with nonconvex un-
certainty sets,” IEEE Transactions on Signal Processing, vol. 71, pp.
218–232, 2023.
\bibitem{grant2011cvx} M. Grant, “CVX: Matlab software for disciplined convex programming,
version 1.21,” http://cvxr. com/cvx, 2011.
\bibitem{9911215} T. Chen, R. Hong, Y. Guo, S. Hao, and B. Hu, “MS²-GNN: Exploring
GNN-based multimodal fusion network for depression detection,” IEEE
Transactions on Cybernetics, vol. 53, no. 12, pp. 7749–7759, 2023.
\bibitem{vaswani2017attention} A. Vaswani, N. Shazeer, N. Parmar, J. Uszkoreit, L. Jones, A. N. Gomez,
Ł. Kaiser, and I. Polosukhin, “Attention is all you need,” Advances in
neural information processing systems, vol. 30, 2017.
\bibitem{10632064} T. Feys, L. V. d. Perre, and F. Rottenberg, “Toward energy-efficient
massive mimo: Graph neural network precoding for mitigating non-
linear PA distortion,” IEEE Transactions on Cognitive Communications
and Networking, vol. 11, no. 1, pp. 184–201, 2025.
\bibitem{xiu2025movable} Y. Xiu, Y. Zhao, R. Yang, D. Niyato, J. Jin, Q. Wang, G. Liu, and N. Wei,
“Movable antenna-aided cooperative ISAC network with time synchro-
nization error and imperfect CSI,” arXiv preprint arXiv:2501.15410,
2025.
\end{thebibliography}
\end{document}